\documentclass[10pt,conference]{IEEEtran}

\IEEEoverridecommandlockouts

\usepackage{amsmath,amssymb,amsfonts}
\usepackage{algorithmic}
\usepackage{graphicx}
\usepackage{textcomp}
\usepackage{xcolor}
\usepackage{ragged2e}
\usepackage{balance}
\usepackage{colortbl}
\usepackage{multirow}
\usepackage{balance}
\usepackage{color}
\usepackage{float}
\usepackage{fancybox}
\usepackage{placeins}
\usepackage{graphicx}

\usepackage{CJKutf8}
\usepackage{indentfirst}
\usepackage{tcolorbox}
\usepackage{makecell}
\tcbuselibrary{skins, breakable, theorems}

\usepackage[colorlinks,
            linkcolor=blue,
            anchorcolor=blue,
            citecolor=blue]{hyperref}
\begin{document}

\title{Understanding Bugs in Multi-Language Deep Learning Frameworks}

\author{
    \IEEEauthorblockN{Zengyang Li\IEEEauthorrefmark{2}\thanks{This work was funded by the Natural Science Foundation of Hubei Province of China under Grant No. 2021CFB577, the National Natural Science Foundation of China under Grant Nos. 62002129 and 62172311, and the Knowledge Innovation Program of Wuhan-Shuguang Project under Grant No. 2022010801020280.}, Sicheng Wang\IEEEauthorrefmark{2}, Wenshuo Wang\IEEEauthorrefmark{2}, Peng Liang\IEEEauthorrefmark{3}\IEEEauthorrefmark{1}\thanks{\IEEEauthorrefmark{1} Corresponding author},  Ran Mo\IEEEauthorrefmark{2}, Bing Li\IEEEauthorrefmark{3}}
    \IEEEauthorblockA{
    \IEEEauthorrefmark{2}School of Computer Science \& Hubei Provincial Key Laboratory of Artificial Intelligence and Smart Learning, \\Central China Normal University, Wuhan, China \\
    \IEEEauthorrefmark{3}School of Computer Science, Wuhan University, Wuhan, China\\
    \{zengyangli, moran\}@ccnu.edu.cn, \{scwang1998, wenshuowang\}@mails.ccnu.edu.cn, \{liangp, bingli\}@whu.edu.cn}
}%end of author

\maketitle

\begin{abstract}
Deep learning frameworks (DLFs) have been playing an increasingly important role in this intelligence age since they act as a basic infrastructure for an increasingly wide range of AI-based applications.
Meanwhile, as multi-programming-language (MPL) software systems, DLFs are inevitably suffering from bugs caused by the use of multiple programming languages (PLs). Hence, it is of paramount significance to understand the bugs (especially the bugs involving multiple PLs, i.e., MPL bugs) of DLFs, which can provide a foundation for preventing, detecting, and resolving bugs in the development of DLFs. 
To this end, we manually analyzed 1497 bugs in three MPL DLFs, namely MXNet, PyTorch, and TensorFlow. First, we classified bugs in these DLFs into 12 types (e.g., \textit{algorithm design bugs} and \textit{memory bugs}) according to their bug labels and characteristics. Second, we further explored the impacts of different bug types on the development of DLFs, and found that \textit{deployment bugs} and \textit{memory bugs} negatively impact the development of DLFs in different aspects the most. Third, we found that 28.6\%, 31.4\%, and 16.0\% of bugs in MXNet, PyTorch, and TensorFlow are MPL bugs, respectively; the PL combination of Python and C/C++ is most used in fixing more than 92\% MPL bugs in all DLFs. Finally, the code change complexity of MPL bug fixes is significantly greater than that of single-programming-language (SPL) bug fixes in all the three DLFs, while in PyTorch MPL bug fixes have longer open time and greater communication complexity than SPL bug fixes. These results provide insights for bug management in DLFs.
%Our analysis of bugs provide insights for bug prevention, detection, and resolution in the future development of DLFs.
\end{abstract}


\section{Introduction}\label{chap:intro}

Deep learning frameworks (DLFs), such as PyTorch, have been playing an increasingly important role in this intelligence age since they act as basic infrastructures for an increasingly wide range of AI-based applications. %The role of DLFs is to reduce the complexity of different use modes of the underlying hardware, provides users with simple application functions, and facilitates users to better use machine resources to complete their customized tasks \cite{mathew2020deep}. 
DLFs provide building blocks for designing, training, and validating deep neural network models through a high-level programming interface. 
Therefore, the reliability of DLFs becomes more and more important for the fast-growing AI-based applications. 
To ensure their reliability, a necessary step is to understand the characteristics and impact of bugs in DLFs. The previous research on bugs in DLFs is mainly divided into two categories: the first category studies the bugs in the implementation of DLFs, e.g., bug categorization, severity, symptoms, root causes, and impacts in various DLFs \cite{r14}\cite{r28}\cite{chen2022toward}\cite{x1}; the second category studies the bugs in the use of specific DLFs, e.g., dependency and performance bugs in deep learning (DL) systems in terms of symptoms, causes, and fix modes \cite{a1}\cite{a2}\cite{r15}\cite{r16}.

Popular DLFs are usually written in multiple programming languages (PLs) \cite{GrEgAd2020}, such as TensorFlow, which is mainly written in Python and C++. Previous research suggests that static code analysis in a multi-programming-language (MPL) software system is much more difficult than in a single-programming-language (SPL) one \cite{shatnawi2019static} \cite{r13}; meanwhile, challenges in understanding multiple PLs and cross-language communication are usually faced by MPL software systems \cite{r12}\cite{r13}. As MPL software systems, DLFs are inevitably suffering from bugs caused by the use of multiple PLs. Therefore, it is of paramount significance to understand the bugs (especially the bugs involving multiple PLs, i.e., MPL bugs) of DLFs, which can provide a foundation for preventing, detecting, and resolving bugs in the development of DLFs. 

%During the research, we found that these popular DLFs are usually written in multiple programming languages (PLs), such as TensorFlow, which is mainly written in Python and C++. At present, software built in multiple PLs has become more and more popular \cite{r2}\cite{r3}\cite{r20}\cite{r47}. According to previous studies, one of the main reasons for using multiple PLs is to reuse existing code with required functions \cite{r5}. Another main reason is to use specific languages to implement certain functions to improve software development efficiency \cite{r6}\cite{r7}\cite{r8}\cite{r9}\cite{r21}. However, the use of multiple PLs will also bring problems to the system. For example, in a multiple-programming-language (MPL) software system, static code analysis is much more difficult than in a single-programming-language (SPL) software system. At the same time, it will also face the problems of understanding multiple PLs and cross-language communication \cite{r12}\cite{r13}. We aim to investigate the \red{causes and selection of multiple PLs in DLFs, as well as the impact on bug fixing}.

In this paper, we investigated the bugs in DLFs. Specifically, we conducted an empirical study on the bugs and their corresponding fixes in three popular DLFs, namely MXNet \cite{r24}, PyTorch \cite{r17} and TensorFlow \cite{r18} on GitHub. The purpose of this work is to systematically understand the bugs and their impacts in the development of DLFs, with a focus on MPL bugs. 
%And from the perspective of DLF as an MPL software system, we explored the causes of MPL and its impact on bug fixing.
%\red{\begin{CJK}{UTF8}{gbsn}[.]\end{CJK}}.
%We finally analyzed 1497 bugs and their corresponding bug-fixing pull requests from MXNet, PyTorch, and TensorFlow on GitHub. We quantitatively and qualitatively analyzed these bugs from the perspectives of \red{DLFs} and MPL software systems. 
The main contributions of this work are threefold:
\begin{itemize}
  \item We conducted an empirical study by manual analysis of 1497 bugs and their corresponding bug fixes from three popular MPL DLFs, namely MXNet, PyTorch, and TensorFlow.  
  \item We classified these bugs based on the labels tagged by developers and bug characteristics, and explored the impacts of bugs on the DLF development in terms of the open time of bugs, the code change complexity of bug fixes, and the communication complexity of developers during bug fixing. 
  \item We explored the MPL bugs and their impact on the three DLFs. Specifically, we looked into the proportion of MPL bugs and the distribution of PL combinations used in each DLF, and the difference between SPL and MPL bugs regarding their impact on the DLFs.
\end{itemize}

The remaining of this paper is organized as follows. Section \ref{chap:relat} presents the related work; Section \ref{chap:case} describes the design of the empirical study; Section \ref{chap:study} presents the results of the study; Section \ref{chap:discus} discusses the study results; Section \ref{chap:threats} identifies the threats to validity of the results; and Section \ref{conclusions} concludes this work with future research directions.

\section{Related Work}\label{RelatedWork}
\label{chap:relat}

\subsection{Bug Classification of DLFs}
\label{RelatedWork_A}
In past years, a number of researchers tried to classify bugs due to different research objectives and perspectives. Islam et al. obtained five types of bugs (e.g., API bugs and structural bugs) in the DLFs through 2716 Stack Overflow posts and 500 bug-fixing commits from GitHub \cite {r16}.
According to the location of the buggy source code, Li et al. obtained a preliminary bug classification of TensorFlow, because TensorFlow tends to place the source files in different directories according to different functions \cite{r25}. Seaman et al. obtained several defect classification schemes applicable to most projects \cite {r39}, then Thung et al. added a new category namely configuration, to the bug classification scheme, and obtained their classification \cite{r14}. Yang et al. summarized the reference architecture of DLFs, based on which they built a bug classification of DLFs \cite{r28}.
Different from the previous classification methods of bugs in DLFs, we employed the grounded theory \cite{r4} and took the bug labels assigned by developers into consideration. Therefore, we obtained a more comprehensive bug classification of DLFs with four newly identified bug types (e.g., deployment and version compatibility bugs). We believe that our bug classification is close to the classification rationale of bug reports by the developers of DLFs, which is convenient for developers to understand which parts of the bugs will be more costly to fix.

\subsection{Impact of the Use of Multiple PLs on Software Systems}\label{RelatedWork_B}
Recently, more and more researchers have begun to pay attention to the impact of the use of multiple PLs on software systems. 
%\cite{r20}\cite{r42}\cite{r43}\cite{r44}\cite{r45}\cite{r46}. 
Ray et al. found a correlation between 11 PLs and software quality in 729 projects hosted on GitHub \cite {r42}. 
%\red{Specifically, their work answers four research questions related to software defects and programming languages.} 
Berger et al. repeated the research of Ray et al. and found that only four PLs were statistically significantly associated with bugs, and the association was very small \cite{r43}. Kochhar et al.  collected a large dataset consisting of 628 projects to study the impact of different PLs on the number of bug fixing commits \cite{r44}. They found that implementing a project with more PLs would increase its defect proneness. Abidi et al. analyzed the source code of the MPL system, and found six anti-patterns \cite{r45} and twelve code smells \cite{r46}. Li et al. analyzed 18 Apache MPL software projects, and confirmed that the use of multiple PLs is related to the increase of development difficulty and the decline of software quality \cite{r20}. Our work further explored the impact of multiple PLs on DLFs.

\section{Study Design}
\label{chap:case}
In order to investigate the bugs in DLFs, we performed an empirical study on mainstream DLFs. In this section, we describe the study, which was designed and reported by following the guidelines proposed by Runeson and H{\"o}st \cite{RuHo2009}.

\subsection{Objective and Research Questions}\label{DesignRQ}
The goal of this study, described using the Goal-Question-Metric (GQM) approach \cite{Ba1992}, is: to \emph{analyze} bugs and their corresponding bug-fixing pull requests \emph{for the purpose of} investigation \emph{with respect to} the bugs with a focus on MPL bugs in DLFs as well as their impacts on DLF development, \emph{from the point of view of} software developers \emph{in the context of} the development of MPL DLFs.

Based on the aforementioned goal, we formulated four research questions (RQs), in which RQ1 and RQ2 focus on the bugs in general in DLFs while RQ3 and RQ4 pay more attention to MPL bugs in DLFs. The RQs are described as follows:

\noindent \textbf{-RQ1}: What is the classification of bugs in DLFs? %What are their \red{symptoms}? 

\noindent \textbf{Rationale}: By classifying bugs in DLFs, we can have a better understanding on the causes and distribution of bugs in DLFs, so that we can conduct an in-depth investigation on each type of the bugs.

\noindent \textbf{-RQ2}: What are the impacts of different types of bugs on the development of DLFs?

\noindent \textbf{Rationale}: Different types of bugs may have different impacts on the development of DLFs. 
%We studied the consequences of these bugs through some indicators we have defined. 
We study the impacts in the aspects of the open time of bugs, the complexity of code change in bug fixing, and the complexity of communication during bug fixing. The answer to this RQ can help recognize the types of bugs that have greatest influence on the development of DLFs, which can be further used to guide the bug management in the development of DLFs.

\noindent \textbf{-RQ3}: What is the proportion of MPL bugs in DLFs? How do MPL bugs distribute over bug types and PL combinations?

\noindent \textbf{Rationale}: By investigating the proportion of MPL bugs in DLFs, we can understand the prevalence of multiple PLs used for bug fixing in DLFs; by looking into the distribution of MPL bugs over bug types and PL combinations, we can get to know the tendency of MPL bugs among bug types and the popularity of different PL combinations used in MPL bug fixing in DLFs.

%\noindent \textbf{-RQ3}: What are the proportions of MPL bug fixes and SPL bug fixes, respectively? What PL combinations are usually used in the MPL bug fixing?

%\noindent \textbf{Rationale}: By investigating the occurrence ratio and language combination of MPL bug fixes compared with SPL bug fixes in DLFs, we can preliminarily understand the prevalence and the use of multiple PLs in DLFs.

%\noindent \textbf{-RQ4}: What are the characteristics of \red{bug type distributions} of MPL fixes and SPL fixes, respectively?

%\noindent \textbf{Rationale}: \red{Analyzing how bugs fixed by MPL and SPL are distributed in types will help us analyze the MPL phenomenon in DLFs.}

\noindent \textbf{-RQ4}: How does the use of multiple PLs affect the bug fixing of DLFs?

\noindent \textbf{Rationale}: With this RQ, we investigate whether the use of multiple PLs can cause additional costs for the bug fixing of DLFs, so as to analyze the impact of the use of multiple PLs on DLFs.

\subsection{Cases and Unit Analysis}\label{CasesandUnitAnalysis}
%According to \cite{RuHo2009}, case studies can be characterized based on the way they define their cases and units of analysis. 
This study investigates DLFs, i.e., cases, and each bug and the corresponding bug-fixing pull request is a single unit of analysis (also called a data unit). 
%In this article, the issue we are studying is marked as a Bug by the project developer. For ease of understanding, we will also refer to these issues as Bugs in the following.

\subsection{Case Selection}\label{CaseSelection}
In this study, we only investigated DLFs hosted on GitHub. The reason we used GitHub projects is that most of DLFs are open source on GitHub. At the same time, this can ensure that all bugs in different DLFs have the same format, so that we can handle the bugs in the same way.
To select each case included in our study, we employed five inclusion criteria:

 \begin{itemize}
  \item C1: The source code written by the main PL is no more than 75\% of the code of the project. This criterion was set to ensure that the PL use is not extremely unbalanced so that the biases caused by specific PLs can be reduced.
  \item C2: It has more than 10k stars, which indicates that it is a popular DLF and has a large group of users.
  \item C3: The DLF has more than 2,000 pull requests. This criterion was set to ensure that the selected DLF is nontrivial, and that the resulting dataset is big enough to be statistically analyzed.
  \item C4: The number of issues of the project is no less than 1,000. This criterion was set to ensure that the selected DLF had been in the maintenance and evolution stage for a reasonable length of period, and thus sufficient data about bug introduction can be collected.
  \item C5: The DLF has more than 150 pairs of bugs and corresponding bug-fixing pull requests. It was set to ensure that the resulting dataset is big enough for statistical analysis.
  \end{itemize}

The thresholds of C1, C3, and C4 were set according to \cite{r20}, and those of C2 and C5 were set based on our experience.

\subsection{Data Collection}\label{DataCollection}
\subsubsection{Data Items to be Collected}
\label{dataitems}
To answer the RQs, we take a bug and its corresponding pull request as the analysis unit, and list the data items to be collected in Table \ref{table:dataitem}. 
For each bug, we need to calculate the open time (OT), lines of source code modified (LOCM), number of files modified (NOFM),
entropy, number of developers participating (NODP), and number
of comments in the pull request (NOC) to quantify the impact of the bug on the development of a DLF. Since most data items are easy to understand, we only explain the definition of entropy in detail \cite{r48}.

Suppose that the source file modified in a pull request for fixing a bug is \{$f_1,f_2,...,f_n$\}. File $f_i(1 \le i \le n)$ is modified $m_i$ times in the pull requests during the last 60 days \cite{LiLiLiMoLi2020} before the current bug-fixing pull request. Let $p_i = m_i / \sum_{j=1}^{n}m_j$. Then, the entropy $H(n) = - \sum_{i=1}^{n}p_i\log_2p_i$.

%\begin{equation}
%p_i = m_i / \sum_{i=1}^{n}m_i .
%\end{equation}

%\noindent Then, the entropy

%\begin{equation}
%H(n) = - \sum_{i=1}^{n}p_i\log_2p_i .
%\end{equation}

\noindent Since the number of modified source files differs between
different bug-fixing pull requests, we need to normalize the entropy to be
comparable. Considering that $H(n)$ achieves the maximum
of $\log_2n$ when $p_i = 1/n  (1 \le i \le n)$, the normalized entropy

\begin{equation}
\tilde{H}(n) =
\begin{cases}
H(n)/log_2n& \text{n \textgreater 1,}\\
0& \text{n = 1.}
\end{cases} 
\end{equation}

The entropy of a bug fix reflects the uncertainty faced by the bug fix. % (i.e., bug-fixing pull request). 
A larger entropy means a bigger uncertainty, resulting in a higher code change complexity.

\begin{table}[h]
\centering
\caption{Data Items To Be Collected.} % 添加标题
\scalebox{0.78}{
\begin{tabular}{|l|l|l|l|}
\hline
\textbf{\#} & \textbf{Name}  & \textbf{Description}                                                                       & \textbf{Relevant RQ} \\ \hline
D1          & IssueID        & The identity number of the issue.                                                          & -                    \\ \hline
D2          & IssueLabels    & The labels of the issue.                                                                   & RQ1, RQ3              \\ \hline
D3          & IssueTitle     & The title of the issue.                                                                    & RQ1                  \\ \hline
D4          & IssueContent   & The description of the issue.                                                              & RQ1                  \\ \hline
D5          & IssueCreatedAt & The creation time of the issue.                                                            & RQ2, RQ4              \\ \hline
D6          & IssueClosedAt  & The close time of the issue.                                                             & RQ2, RQ4              \\ \hline
D7          & PrID           & \makecell[l]{The identity number of the pull request \\associated with the issue.}                         & -                    \\ \hline
D8          & PrLabels       & The labels of the pull request.                                                            & RQ1, RQ3              \\ \hline
D9         & PrTitle        & The title of the pull request.                                                             & RQ1                  \\ \hline
D10         & PrContent      & The description of the pull request.                                                       & RQ1                  \\ \hline
D11         & Files          & The files modified in the pull request.                                                    & RQ2- 
 RQ4              \\ \hline
D12         & OT             & The open time of the issue, i.e., OT=D6-D5. & RQ2, RQ4              \\ \hline
D13         & LOCM           & \makecell[l]{The number of lines of source code modified \\in the pull request.}                           & RQ2, RQ4              \\ \hline
D14         & NOFM           & The number of files modified in the pull request.                                          & RQ2, RQ4              \\ \hline
D15         & Entropy        & \makecell[l]{The normalized entropy of the modified files \\in the pull request.}                          & RQ2, RQ4              \\ \hline
D16         & NODP           & \makecell[l]{The number of developers participating in the \\pull request (excluding robots).}                   & RQ2, RQ4              \\ \hline
D17         & NOC            & \makecell[l]{The number of comments in the pull request \\(excluding robots).}                             & RQ2, RQ4              \\ \hline
D18         & IsMPLF         & \makecell[l]{Whether the pull request is a MPL fix, i.e., \\involving source files in multiple PLs.}                                                     & RQ3, RQ4              \\ \hline

\end{tabular}
}
\label{table:dataitem}
\end{table}

\subsubsection{Data Collection Procedure}\label{datacollectionprocedure}
To collect the data items in Table \ref{table:dataitem}, we developed a dedicated tool based on GitHub GraphQL APIs to extract the required data from the repository of each DLF on GitHub. Specifically, we collected the data in the following steps:

%The specific filtering process includes the following steps (shown in Fig.\ref{fig:Procedure}).

\noindent \textbf{Step 1}: Export available issue reports. Extract all related issues and corresponding pull requests in the project. 
%Such a report is a complete problem fix process.

\noindent \textbf{Step 2}: Filter bug reports. The filtering logic (satisfying anyone below) is that: an issue or the pull request connected to it has at least one whose label's name or label's description involving the word "Bug"; the word "Bug" or "bug" or "BUG" appears in the issue title; issue description adopts a bug report template. In this way, we can get the bug reports we need from all available reports.

\noindent \textbf{Step 3}: Filter available bugs. The status of the issue must be ``closed'' and the status of the pull request corresponding to the issue must be ``merged''. This ensures that the bug has an effective fix.

%\noindent \textbf{Step 4}: A project that ensures sufficient experimental data. We need more than 200 experimental data, which is enough for us to carry out simple data statistics. Through this step, we get MXNet, PyTorch, and TensorFlow. 

\noindent \textbf{Step 4}: Remove abnormal data. An abnormal data unit means that a bug does not strictly correspond to a single bug-fixing pull request (e.g., a pull request corresponds to multiple bugs, a bug corresponds to multiple pull requests). This will lead to inaccurate results of the
impact of the bug on the project.

\noindent \textbf{Step 5}: Calculate the data items listed in Table \ref{table:dataitem}.

%Through the data filtering process, we got MXNet, PyTorch, and TensorFlow, which are the most popular DLFs. The main reason for removing a large amount of data is that only a few issues are bound to pull requests. However, due to the similarity of DLFs and the similar language combination they adopt. Therefore, we believe that the research on these three frameworks can also be used to characterize other DLFs.

%We collected 1909 bugs and their corresponding fixes from three DLFs. To get more accurate results, we manually removed some abnormal data (exceptions here refer to the following four situations: one pull request corresponds to multiple issues; one issue corresponds to multiple pull requests; pull request only solves a small part of issues; pull request is a repair stack. We think these four exceptions will affect our results.). Through checking, 1499 pieces of data were finally obtained for our research.

%So far, we have completed the collection of experimental data.

%\begin{figure}[htp]
%    \centering
%    \includegraphics[width=3cm]{Images/procedure.png}
%    \caption{Procedure of data collection.}
%    \label{fig:Procedure}
%\end{figure}

\subsection{Data Analysis}

\subsubsection{Data Analysis for RQ1}\label{DataAnalysisRQ1}

For RQ1, to obtain the bug classification in DLFs, we followed a four-step process below.\\
\noindent \textbf{Step 1: Preliminarily classify bugs based on bug labels.} This step is to get a preliminary bug classification. (1) We collected all the labels of bugs in the selected DLFs through a dedicated tool. (2) We then examined relevant official documents and label descriptions
%, and \blue{the description} 
of bugs to have a deep understanding of the labels. (3) Based on the bug labels and characteristics, 
%the first three authors \red{discussed XXX}
the second and third authors classify bug labels separately. Then, the first author reviewed the label classification and solved disagreements through discussion with the second and third authors. Finally, we obtained a preliminary classification of the bugs in DLFs. 

\noindent \textbf{Step 2: Construct a relatively stable bug classification.} The first two authors manually analyzed a set of sampled bugs with the help of grounded theory \cite{r4}. When the classification obtained in \textbf{Step 1} was found inappropriate during the bug analysis, the first two authors tried to improve the bug classification through discussion. In this step, new bug types might arise, existing bug types might be removed, and the bug classification would be updated accordingly. Finally, a relatively stable classification was obtained. 

\noindent \textbf{Step 3: Conduct a pilot bug tagging.} This step is to get a stable bug classification and reach a common understanding on each bug type in the bug classification. 
%\red{We manually analyzed bugs with the help of grounded theory \cite{r4}}. 
We randomly selected 10\% of the bugs, and the second and third authors separately tagged each of the bugs with an appropriate bug type from the obtained classification got in \textbf{Step 1}. If there was any disagreement on bug tagging, the second and third authors discussed them with the first author to get a consensus.    
%If the original classification is inappropriate in the process of bug analysis, the first two authors will discuss resolving the conflict and modifying the classification. Finally, a more stable classification result is obtained. 
We used Cohen's Kappa \cite{r23} to measure the consistency between the bug tagging results of the two authors. When the Cohen's Kappa value is less than 90\%, the first three authors needed to discuss for resolving disagreements and randomly extracted another 10\% of the bugs for another round of bug tagging. The bug tagging was an iterative process, which stopped when the Cohen's Kappa value exceeded 90\%. 

\noindent \textbf{Step 4: Classify the remaining bugs.} The second author classified the remaining bugs into different bug types.

\subsubsection{Data Analysis for RQ2-RQ4}
%For RQ2-RQ4, only descriptive statistics are used. 
To answer RQ2, we investigated the impact of different types of bugs on DLF development through six impact indicators (i.e., data items D12-D17) in three aspects: (1) open time of bugs, i.e., OT (D12); (2) complexity of code change in bug fixing, including LOCM (D13), NOFM (D14), and entropy (D15); and (3) complexity of communication during bug fixing, including NODP (D17) and NOC (D18). For each indicator, we calculated the average value and ranking of each bug type in each DLF, and also calculated the mean ranking for all the selected DLFs by averaging their ranking numbers. The integrated ranking (InteRanking) numbers for all bug types are assigned according to their mean rankings.
To answer RQ3, we examined the extensions of the modified source files in the bug-fixing pull requests to identify the MPL and SPL fixes, and calculated the bug type distribution of bugs with these MPL and SPL fixes. In addition, we calculated the distribution of MPL bugs over different PL combinations. Similar as the PLs examined in \cite{r20}, we only considered the following 18 general-purpose PLs: C/C++, C\#, Clojure, CoffeeScript, Erlang, Go, Haskell, Java, JavaScript, Kotlin, Objective-C, Perl, PHP, Python, Ruby, Scala, Swift, and TypeScript. 
%listed in Table \ref{table2}. %\red{It means that script PLs and configuration documents were excluded to increase the accuracy of the study results.}
To answer RQ4, we studied the difference between MPL bug fixes and SPL bug fixes regarding their impact on DLF development through Mann-Whitney U tests. Since the data of the variables to be tested do not necessarily follow a specific distribution, it is reasonable to use the Mann-Whitney U test – a non-parametric test – in this study. The test was significant at \textit{p-value}\textless{}0.05, which means that the tested groups have a significant difference. 

%\begin{table}[h]
%\Central
%\caption{Programming Languages Examined.}
%\begin{center} 
%\begin{tabular}{|l|l|l|l|l|l|}
%\hline
%\textbf{\#} & \textbf{PL}           & \textbf{\#} & \textbf{PL}          & \textbf{\#} & \textbf{PL}         \\ \hline
%1  & C/C++        & 7  & Haskell     & 13 & PHP        \\ \hline
%2  & C\#          & 8  & Java        & 14 & Python     \\ \hline
%3  & Clojure      & 9  & JavaScript  & 15 & Ruby       \\ \hline
%4  & CoffeeScript & 10 & Kotlin      & 16 & Scala      \\ \hline
%5  & Erlang       & 11 & Objective-C & 17 & Swift      \\ \hline
%6  & Go           & 12 & Perl        & 18 & TypeScript \\\hline
%\end{tabular}
%\end{center}
%\label{table2}
%\end{table}

\section{Study Results}
\label{chap:study}

%\red{Our initial selection of datasets includes some popular DLFs on GitHub, including  Caffe, Chainer, Deeplearning4J, Keras, MXNet, PaddlePaddle, PyTorch, TensorFlow, TFLearn, and Theano. We made a preliminary inspection of these projects and found that not all the DLFs are suitable for our experiment. \blue{The main reason is that only a few issues are bound to pull requests. This will lead to insufficient experimental data for our research.} }
Through case selection, we got three DLFs, i.e., MXNet, PyTorch, and TensorFlow, which demographic information is shown in Table \ref{table:demographicInfo}. To be specific, 189, 926, and 382 data units for analysis were collected from the whole bug set of MXNet, PyTorch, and TensorFlow respectively, and 1497 data units in total were obtained for investigation. The dataset is available online \cite{DatasetsICPC2022}. Some DLFs such as Caffe and Keras were not included, since only a few bugs are bound to their bug-fixing pull requests or only a single general-purpose PL is used.

\begin{table}[h]
\centering
\caption{Demographic Information of The Selected DLFs.}
\scalebox{0.85}{
\begin{tabular}{|c|c|c|c|c|c|}
\hline
\textbf{DLF}           & \textbf{\#Pr}  & \textbf{\#Issue} & \textbf{\#star (k)} & \textbf{\%Main PL} & \textbf{\#Unit} \\ \hline
MXNet      & 11096 & 9532    & 20.2      & 48.6      & 189    \\ \hline
PyTorch    & 59555 & 29575   & 60.3      & 49.4      & 926    \\ \hline
TensorFlow & 22191 & 36007   & 169.0     & 63.1      & 382    \\ \hline
Total      & 92842 & 75114   & 289.5     & -         & 1497   \\ \hline
\end{tabular}
}
\label{table:demographicInfo}
\end{table}

\subsection{Bug Types in Deep Learning Frameworks (RQ1)}
%In the first round of pilot bug tagging, the Kappa value is 80\%. After two rounds of analysis, the kappa value exceeds 90\%. 

\begin{figure*}[]
    \centering
    \includegraphics[width=16cm]{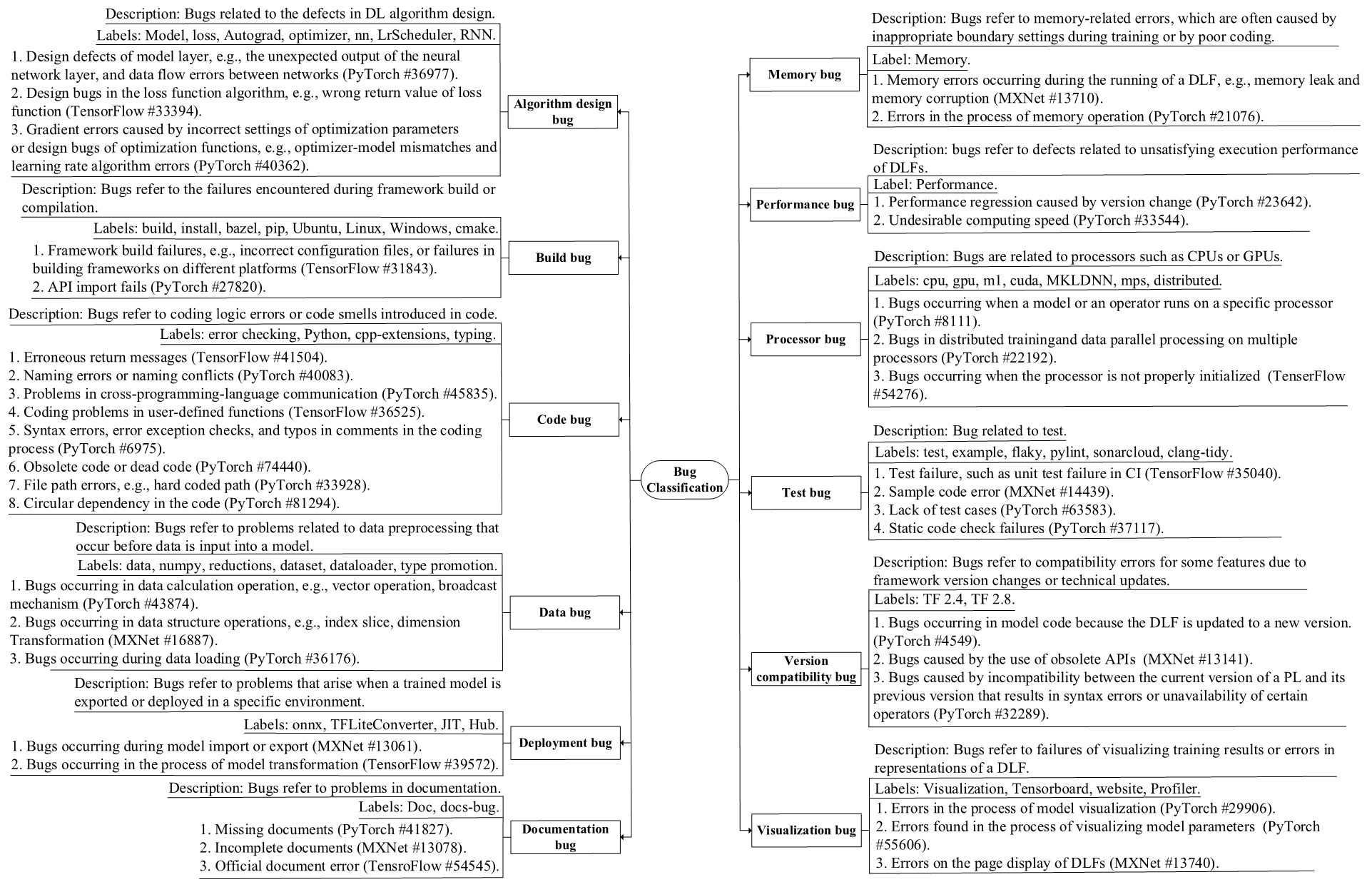}
    \caption{Bug types in DLFs (RQ1).}
    \label{fig:BugType1}
\end{figure*}

Through manual analysis following the process presented in Section \ref{DataAnalysisRQ1}, we classified the collected 1497 bugs in the three DLFs into 12 types, which are shown in Fig.~\ref{fig:BugType1}. The details of the 12 bug types are described as follows.

\begin{table}[h]
\centering
\caption{Numbers and Proportions of Different Bug Types (RQ1).}
\scalebox{0.84}{
\begin{tabular}{|l|rr|rr|rr|rr|}
\hline
\multicolumn{1}{|c|}{\multirow{2}{*}{\textbf{Bug Type}}} & \multicolumn{2}{c|}{\textbf{MXNet}}            & \multicolumn{2}{c|}{\textbf{PyTorch}}          & \multicolumn{2}{c|}{\textbf{TensorFlow}}       & \multicolumn{2}{c|}{\textbf{Total}}            \\ \cline{2-9} 
\multicolumn{1}{|c|}{}                                   & \multicolumn{1}{c|}{\textbf{\#}} & \textbf{\%} & \multicolumn{1}{c|}{\textbf{\#}} & \textbf{\%} & \multicolumn{1}{c|}{\textbf{\#}} & \textbf{\%} & \multicolumn{1}{c|}{\textbf{\#}} & \textbf{\%} \\ \hline
Algorithm design bug                                     & \multicolumn{1}{c|}{21}          & 11.1       & \multicolumn{1}{c|}{119}         & 12.9       & \multicolumn{1}{c|}{34}          & 8.9        & \multicolumn{1}{c|}{174}         & 11.6       \\ \hline
Build bug                                                & \multicolumn{1}{c|}{32}          & 16.9       & \multicolumn{1}{c|}{112}         & 12.1       & \multicolumn{1}{c|}{31}          & 8.1        & \multicolumn{1}{c|}{175}         & 11.7       \\ \hline
Code bug                                                 & \multicolumn{1}{c|}{15}          & 7.9        & \multicolumn{1}{c|}{81}          & 8.8        & \multicolumn{1}{c|}{45}          & 11.8       & \multicolumn{1}{c|}{141}         & 9.4        \\ \hline
Data bug                                                 & \multicolumn{1}{c|}{45}          & 23.8       & \multicolumn{1}{c|}{155}         & 16.7       & \multicolumn{1}{c|}{96}          & 25.1       & \multicolumn{1}{c|}{296}         & 19.8       \\ \hline
Deployment bug                                           & \multicolumn{1}{c|}{7}           & 3.7        & \multicolumn{1}{c|}{106}         & 11.5       & \multicolumn{1}{c|}{25}          & 6.5        & \multicolumn{1}{c|}{138}         & 9.2        \\ \hline
Documentation bug                                        & \multicolumn{1}{c|}{15}          & 7.9        & \multicolumn{1}{c|}{29}          & 3.1        & \multicolumn{1}{c|}{80}          & 20.9       & \multicolumn{1}{c|}{124}         & 8.3        \\ \hline
Memory bug                                               & \multicolumn{1}{c|}{4}           & 2.1        & \multicolumn{1}{c|}{27}          & 2.9        & \multicolumn{1}{c|}{6}           & 1.6        & \multicolumn{1}{c|}{37}          & 2.5        \\ \hline
Performance bug                                          & \multicolumn{1}{c|}{2}           & 1.1        & \multicolumn{1}{c|}{22}          & 2.4        & \multicolumn{1}{c|}{0}           & 0.0        & \multicolumn{1}{c|}{24}          & 1.6        \\ \hline
Processor bug                                            & \multicolumn{1}{c|}{10}          & 5.3        & \multicolumn{1}{c|}{132}         & 14.3       & \multicolumn{1}{c|}{13}          & 3.4        & \multicolumn{1}{c|}{155}         & 10.4       \\ \hline
Test bug                                                 & \multicolumn{1}{c|}{19}          & 10.1       & \multicolumn{1}{c|}{83}          & 9.0       & \multicolumn{1}{c|}{23}          & 6.0        & \multicolumn{1}{c|}{125}         & 8.4        \\ \hline
Version compatibility bug                                & \multicolumn{1}{c|}{10}          & 5.3        & \multicolumn{1}{c|}{48}          & 5.2        & \multicolumn{1}{c|}{18}          & 4.7        & \multicolumn{1}{c|}{76}          & 5.1        \\ \hline
Visualization bug                                        & \multicolumn{1}{c|}{9}           & 4.8        & \multicolumn{1}{c|}{12}          & 1.3        & \multicolumn{1}{c|}{11}          & 2.9        & \multicolumn{1}{c|}{32}          & 2.1        \\ \hline
\end{tabular}
}
\label{table:bugtypes}
\end{table}

\textbf{(1) Algorithm design bug.} This type of bugs are related to the defects in DL algorithm design. There are three specific cases: 
1) design bugs in the model layer, e.g., 
%the output beyond the expectation of the neural network layer, 
network structure errors, model accuracy deterioration, and data flow errors between networks;
2) design bugs in the loss function algorithm, e.g., the incorrect return value of the loss function and the internal structure error of the loss function; 
3) gradient errors caused by incorrect settings of optimization parameters or design bugs of optimization functions, e.g., optimizer-model mismatches and learning rate algorithm errors.

\textbf{(2) Build bug.} This type of bugs refer to the failures encountered during framework build or compilation. There are two specific cases: 1) framework build failures, which may be caused by configuration file errors or by failures when building frameworks in different platforms, such as Linux, Windows, Mac, or Docker environments; and 2) API import failures. 
%3)errors in different compilation environments (\textit{PyTorch \#37584}).     

\textbf{(3) Code bug.} This type of bugs refer to coding logic errors or code smells introduced in code. Typical instances include 1) erroneous return messages; 2) naming errors or naming conflicts; 3) problems in cross-programming-language communication; 4) coding problems in user-defined functions, e.g., design defects of class templates; 5) obsolete code or dead code; 6) file path errors, e.g., hard coded path and path identification errors; and 7) circular dependencies.

\textbf{(4) Data bug.} This type of bugs refer to problems related to data preprocessing that occur before data is input into a model. Their symptoms are usually function output errors, data loading errors, data acquisition exceptions, data corruption, parameter errors, etc. There are three specific cases: 1) bugs occurring in data calculation operations, e.g., scalar operations, vector operations, matrix operations, and broadcast mechanisms; 2) bugs occurring in data structure operations, e.g., data creation, data replication, index slicing, type conversion, and other related errors; and 3) bugs occurring during data loading.

\textbf{(5) Deployment bug.} This type of bugs refer to problems that arise when a trained model is exported or deployed in a specific environment. There are two cases: 1) bugs occurring during model import or export, e.g., abnormal behaviors when storing trained models; and 2) bugs occurring in the process of model transformation.

\textbf{(6) Documentation bug.} This type of bugs refer to problems in documentation. Typical bugs: 1) missing documents; 2) incomplete documents; and 3) official document errors.

\textbf{(7) Memory bug.} This type of bugs refer to memory-related errors, which are often caused by inappropriate boundary settings during training or by poor coding. There are two main cases: 1) memory errors occurring during the running of a DLF, e.g., memory leak, memory corruption, illegal memory access, and insufficient memory; and 2) errors in the process of memory operation.

\textbf{(8) Performance bug.} This type of bugs refer to defects related to unsatisfying execution performance of DLFs. Typical bugs include: 1) undesirable computing speed; and 2) performance regression caused by version change.
%during the development process, developers tend to tag some bugs with Performance labels, although some of them do not have the impact of bugs. But they are still marked as bugs, so we hope to discuss them. They are bugs raised separately to solve performance problems, such as undesirable computing speed and performance regression caused by version change.

\textbf{(9) Processor bug.} This type of bugs are related to processors, such as CPUs or GPUs. There are three cases: 1) bugs occurring when a model or an operator runs on a specific processor; 2) bugs in distributed training and data parallel processing on multiple processors; and 3) bugs occurring when the processor is not properly initialized, e.g., the processor does not match some platform environments.

\textbf{(10) Test bug.} This type of bugs refer to problems in testing, including 1) test failures; 2) sample code errors; 3) lack of test cases; and 4) static code check failures.

\textbf{(11) Version compatibility bug.} This type of bugs refer to compatibility errors for some features due to framework version changes or technical updates. There are three cases: 1) Bugs occurring in model code because the DL framework is updated to a new version; 2) bugs caused by the use of obsolete APIs; and 3) bugs caused by incompatibility between the current version of a PL and its previous version that results in syntax errors or unavailability of certain operators.

\textbf{(12) Visualization bug.} This type of bugs refer to failures of visualizing training results or errors in representations of a DLF. There are three cases: 1) errors in the process of model visualization; 2) errors found in the process of visualizing model parameters; 3) errors on the page display of DLFs, e.g., broken links at the front and page display errors.

The number and proportion of each bug type are presented in TABLE \ref{table:bugtypes}, which shows that the proportions of the bug types are not well balanced either in individual DLFs or in all of them as a whole. In MXNet, data bugs are with the largest proportion (23.8\%) and performance bugs have the least proportion (1.1\%). In PyTorch, data bugs are with the largest proportion (16.7\%) and visualization bugs have the least proportion (1.3\%). In TensorFlow, data bugs are with the largest proportion (25.1\%) and there are no performance bugs in our collected dataset. Taking all DLFs as a whole, data bugs (19.8\%) are the leading bug type, followed by build bugs (11.7\%) and algorithm design bugs (11.6\%); the least bug types are performance bugs (1.6\%), followed by visualization bugs (2.1\%) and memory bugs (2.5\%). 

\textbf{Answer to RQ1:} Bugs in DLFs can be classified into 12 types. \textit{Data bugs} are the dominant bug type (19.8\%) and \textit{performance bugs} account for the smallest proportion (1.6\%).

%\begin{tcolorbox}[colback=white!5]
%  \textbf{Answer to RQ1:} Bugs in DLFs can be classified into 12 types: algrithm design bug, build bug, code bug, data bug, deployment bug, documentation bug, memory bug, performance bug, test bug, version compatibility bug, and visualization bug.
%\end{tcolorbox}

\subsection{Impacts of Bugs on DLF Development (RQ2)}

We present the impacts of different types of bugs on the development of DLFs in three aspects, i.e., the open time of bugs, the complexity of code change in bug fixing, and the complexity of communication during bug fixing. 
%\red{We take the result presentation method used in \cite{r14} as a reference, in order to reduce the inaccuracy caused by fewer data and find out the common impact of bugs on the three frameworks as much as possible.}

\subsubsection{Open Time of Bugs}
We compared the associations between bug and open time in the three DLFs in different categories horizontally, as shown in Table \ref{table3}, in which InteRanking denotes the integrated ranking of an indicator according to its mean ranking. For MXNet, visualization bugs and performance bugs are ranked in the top and bottom respectively; for PyTorch, performance bugs and version compatibility bugs are ranked in the top and bottom respectively; and for TensorFlow, memory bugs and test bugs are ranked in the top and bottom respectively. Taking the three DLFs as a whole, deployment bugs, documentation bugs, and memory bugs are the top 3 bug types that cost the most OT during bug fixing; build bugs, test bugs, and processor bugs cost the least OT during bug fixing.
% Taking all DLFs as a whole, most bugs are fixed within one month (i.e., 30 days) after they are reported, while some bugs take longer to fix. \red{To get a better understanding of what types of bugs tend to take longer to fix, we mark the bug types whose proportion of open time more than a month exceeds 40%.} 
%Although we will inevitably be affected by different project development habits and team size, we can still get some information we want from the data. 
% \red{For MXNet, deployment bugs, memory bugs, and version compatibility bugs show a relatively long open time; for PyTorch, algorithm design bugs, deployment bugs, memory bugs, and performance bugs take longer to fix; for TensorFlow, there are algorithm design bugs, deployment bugs, memory bugs, and processor bugs. }

% \textbf{From the common point of three DLFs, deployment bugs and memory bugs have longer open time, followed by algorithm design bugs, which also have a strong impact on two of the three frameworks.}

\begin{table*}[h]
\centering
\caption{Open Time (in days) of Different Bug Types (RQ2).} 
\scalebox{0.85}{
\begin{tabular}{|c|cc|cc|cc|cc|}
\hline
\multicolumn{1}{|l|}{}    & \multicolumn{2}{c|}{\textbf{MXNet}}                 & \multicolumn{2}{c|}{\textbf{PyTorch}}               & \multicolumn{2}{c|}{\textbf{TensorFlow}}            & \multicolumn{2}{c|}{\textbf{Total}}                           \\ \hline
\textbf{Bug type}         & \multicolumn{1}{c|}{\textbf{OT}} & \textbf{Ranking} & \multicolumn{1}{c|}{\textbf{OT}} & \textbf{Ranking} & \multicolumn{1}{c|}{\textbf{OT}} & \textbf{Ranking} & \multicolumn{1}{c|}{\textbf{Mean ranking}} & \textbf{InteRanking} \\ \hline
Algorithm design bug      & \multicolumn{1}{c|}{22.92}       & 10               & \multicolumn{1}{c|}{74.70}       & 3                & \multicolumn{1}{c|}{130.32}      & 2                & \multicolumn{1}{c|}{5.00}                  & 4                \\ \hline
Build bug                 & \multicolumn{1}{c|}{40.24}       & 7                & \multicolumn{1}{c|}{30.63}       & 10               & \multicolumn{1}{c|}{33.86}       & 10               & \multicolumn{1}{c|}{9.00}                  & 11=              \\ \hline
Code bug                  & \multicolumn{1}{c|}{46.34}       & 6                & \multicolumn{1}{c|}{57.68}       & 5                & \multicolumn{1}{c|}{47.30}       & 7                & \multicolumn{1}{c|}{6.00}                  & 5=               \\ \hline
Data bug                  & \multicolumn{1}{c|}{55.23}       & 5                & \multicolumn{1}{c|}{39.63}       & 9                & \multicolumn{1}{c|}{68.15}       & 5                & \multicolumn{1}{c|}{6.33}                  & 7                \\ \hline
Deployment bug            & \multicolumn{1}{c|}{77.84}       & 3                & \multicolumn{1}{c|}{81.51}       & 2                & \multicolumn{1}{c|}{66.53}       & 6                & \multicolumn{1}{c|}{3.67}                  & 1                \\ \hline
Documentation bug         & \multicolumn{1}{c|}{81.32}       & 2                & \multicolumn{1}{c|}{50.53}       & 6                & \multicolumn{1}{c|}{84.38}       & 4                & \multicolumn{1}{c|}{4.00}                  & 2                \\ \hline
Memory bug                & \multicolumn{1}{c|}{40.18}       & 8                & \multicolumn{1}{c|}{68.90}       & 4                & \multicolumn{1}{c|}{135.39}      & 1                & \multicolumn{1}{c|}{4.33}                  & 3                \\ \hline
Performance bug           & \multicolumn{1}{c|}{0.75}        & 12               & \multicolumn{1}{c|}{106.09}      & 1                & \multicolumn{1}{c|}{-}           & -                & \multicolumn{1}{c|}{6.50}                  & 8                \\ \hline
Processor bug             & \multicolumn{1}{c|}{21.41}       & 11               & \multicolumn{1}{c|}{30.23}       & 11               & \multicolumn{1}{c|}{99.04}       & 3                & \multicolumn{1}{c|}{8.33}                  & 10               \\ \hline
Test bug                  & \multicolumn{1}{c|}{27.45}       & 9                & \multicolumn{1}{c|}{48.62}       & 7                & \multicolumn{1}{c|}{32.07}       & 11               & \multicolumn{1}{c|}{9.00}                  & 11=              \\ \hline
Version compatibility bug & \multicolumn{1}{c|}{58.45}       & 4                & \multicolumn{1}{c|}{17.09}       & 12               & \multicolumn{1}{c|}{38.47}       & 8                & \multicolumn{1}{c|}{8.00}                  & 9                \\ \hline
Visualization bug         & \multicolumn{1}{c|}{82.84}       & 1                & \multicolumn{1}{c|}{43.89}       & 8                & \multicolumn{1}{c|}{37.17}       & 9                & \multicolumn{1}{c|}{6.00}                  & 5=               \\ \hline
\end{tabular}
}
\label{table3}
\end{table*}

\subsubsection{Complexity of Code Change in Bug Fixing}
We investigated the impact of different types of bugs on the complexity of code change in terms of LOCM, NOFM, and entropy of bug-fixing pull requests in the three selected DLFs.

\textbf{(1) LOCM.} The average LOCM and its ranking of each bug type in the three DLFs are shown in Table \ref{table:LOCM}. For MXNet, memory bugs and performance bugs are ranked in the top and bottom respectively; for PyTorch, deployment bugs and documentation bugs are ranked in the top and bottom respectively; and for TensorFlow, processor bugs and build bugs are ranked in the top and bottom respectively. 
Taking the three DLFs as a whole, memory bugs, algorithm design bugs, code bugs, and deployment bugs are the top 4 bug types that need the most LOCM during bug fixing; build bugs, documentation bugs, and visualization bugs need the least LOCM during bug fixing.

\begin{table*}[h]
\caption{Average LOCM and Its Ranking of Each Bug Type in DLFs (RQ2).}
\centering
\scalebox{0.85}{
\begin{tabular}{|c|cc|cc|cc|cc|}
\hline
\multicolumn{1}{|l|}{}    & \multicolumn{2}{c|}{\textbf{MXNet}}                   & \multicolumn{2}{c|}{\textbf{PyTorch}}                 & \multicolumn{2}{c|}{\textbf{TensorFlow}}              & \multicolumn{2}{c|}{\textbf{Total}}                           \\ \hline
\textbf{Bug type}         & \multicolumn{1}{c|}{\textbf{LOCM}} & \textbf{Ranking} & \multicolumn{1}{c|}{\textbf{LOCM}} & \textbf{Ranking} & \multicolumn{1}{c|}{\textbf{LOCM}} & \textbf{Ranking} & \multicolumn{1}{c|}{\textbf{Mean ranking}} & \textbf{InteRanking} \\ \hline
Algorithm design bug      & \multicolumn{1}{c|}{66.10}         & 2                & \multicolumn{1}{c|}{60.71}         & 2                & \multicolumn{1}{c|}{32.50}         & 4                & \multicolumn{1}{c|}{2.67}                  & 2                \\ \hline
Build bug                 & \multicolumn{1}{c|}{34.72}         & 7                & \multicolumn{1}{c|}{26.22}         & 10               & \multicolumn{1}{c|}{14.58}         & 11               & \multicolumn{1}{c|}{9.33}                  & 12               \\ \hline
Code bug                  & \multicolumn{1}{c|}{39.20}         & 6                & \multicolumn{1}{c|}{54.91}         & 4                & \multicolumn{1}{c|}{30.96}         & 5                & \multicolumn{1}{c|}{5.00}                  & 3=                \\ \hline
Data bug                  & \multicolumn{1}{c|}{34.27}         & 8                & \multicolumn{1}{c|}{46.19}         & 7                & \multicolumn{1}{c|}{29.59}         & 6                & \multicolumn{1}{c|}{7.00}                  & 7                \\ \hline
Deployment bug            & \multicolumn{1}{c|}{28.14}         & 11               & \multicolumn{1}{c|}{61.97}         & 1                & \multicolumn{1}{c|}{51.68}         & 3                & \multicolumn{1}{c|}{5.00}                  & 3=                \\ \hline
Documentation bug         & \multicolumn{1}{c|}{43.93}         & 4                & \multicolumn{1}{c|}{18.79}         & 12               & \multicolumn{1}{c|}{15.78}         & 10               & \multicolumn{1}{c|}{8.67}                  & 10=                \\ \hline
Memory bug                & \multicolumn{1}{c|}{87.75}         & 1                & \multicolumn{1}{c|}{55.33}         & 3                & \multicolumn{1}{c|}{57.17}         & 2                & \multicolumn{1}{c|}{2.00}                  & 1                \\ \hline
Performance bug           & \multicolumn{1}{c|}{15.00}         & 12               & \multicolumn{1}{c|}{54.05}         & 5                & \multicolumn{1}{c|}{-}             & -                & \multicolumn{1}{c|}{8.50}                  & 9                \\ \hline
Processor bug             & \multicolumn{1}{c|}{28.50}         & 10               & \multicolumn{1}{c|}{51.52}         & 6                & \multicolumn{1}{c|}{57.69}         & 1                & \multicolumn{1}{c|}{5.67}                  & 5                \\ \hline
Test bug                  & \multicolumn{1}{c|}{39.26}         & 5                & \multicolumn{1}{c|}{23.75}         & 11               & \multicolumn{1}{c|}{18.52}         & 8                & \multicolumn{1}{c|}{8.00}                  & 8                \\ \hline
Version compatibility bug & \multicolumn{1}{c|}{58.30}         & 3                & \multicolumn{1}{c|}{26.31}         & 9                & \multicolumn{1}{c|}{25.83}         & 7                & \multicolumn{1}{c|}{6.33}                  & 6                \\ \hline
Visualization bug         & \multicolumn{1}{c|}{31.33}         & 9                & \multicolumn{1}{c|}{45.75}         & 8                & \multicolumn{1}{c|}{16.18}         & 9                & \multicolumn{1}{c|}{8.67}                  & 10=                \\ \hline
\end{tabular}
}
\label{table:LOCM}
\end{table*}

\textbf{(2) NOFM.} The average NOFM and its ranking of each bug type in the three DLFs are shown in Table \ref{table:NOFM}. For MXNet, memory bugs and performance bugs are ranked in the top and bottom respectively; for PyTorch, deployment bugs and documentation bugs are ranked in the top and bottom respectively; and for TensorFlow, version compatibility bugs and documentation bugs are ranked in the top and bottom respectively. Taking the three DLFs as a whole, memory bugs, processor bugs, and algorithm design bugs are the top 3 bug types that need the most NOFM during bug fixing; performance bugs, documentation bugs, test bugs, and visualization bugs need the least NOFM during bug fixing.

\begin{table*}[h]
\centering
\caption{Average NOFM and Its Ranking of Each Bug Type in DLFs  (RQ2).} %
\scalebox{0.85}{
\begin{tabular}{|c|cc|cc|cc|cc|}
\hline
\multicolumn{1}{|l|}{}    & \multicolumn{2}{c|}{\textbf{MXNet}}                   & \multicolumn{2}{c|}{\textbf{PyTorch}}                 & \multicolumn{2}{c|}{\textbf{TensorFlow}}              & \multicolumn{2}{c|}{\textbf{Total}}                           \\ \hline
\textbf{Bug type}         & \multicolumn{1}{c|}{\textbf{NOFM}} & \textbf{Ranking} & \multicolumn{1}{c|}{\textbf{NOFM}} & \textbf{Ranking} & \multicolumn{1}{c|}{\textbf{NOFM}} & \textbf{Ranking} & \multicolumn{1}{c|}{\textbf{Mean ranking}} & \textbf{InteRanking} \\ \hline
Algorithm design bug      & \multicolumn{1}{c|}{3.00}          & 5                & \multicolumn{1}{c|}{2.99}          & 3                & \multicolumn{1}{c|}{1.97}          & 9                & \multicolumn{1}{c|}{5.67}                  & 3                \\ \hline
Build bug                 & \multicolumn{1}{c|}{3.03}          & 4                & \multicolumn{1}{c|}{1.99}          & 9                & \multicolumn{1}{c|}{2.25}          & 5                & \multicolumn{1}{c|}{6.00}                  & 4                \\ \hline
Code bug                  & \multicolumn{1}{c|}{2.73}          & 7                & \multicolumn{1}{c|}{2.80}          & 5                & \multicolumn{1}{c|}{2.04}          & 7                & \multicolumn{1}{c|}{6.33}                  & 5                \\ \hline
Data bug                  & \multicolumn{1}{c|}{2.31}          & 10               & \multicolumn{1}{c|}{2.87}          & 4                & \multicolumn{1}{c|}{2.18}          & 6                & \multicolumn{1}{c|}{6.67}                  & 6=                \\ \hline
Deployment bug            & \multicolumn{1}{c|}{2.57}          & 9                & \multicolumn{1}{c|}{3.25}          & 1                & \multicolumn{1}{c|}{1.96}          & 10               & \multicolumn{1}{c|}{6.67}                  & 6=                \\ \hline
Documentation bug         & \multicolumn{1}{c|}{3.53}          & 2                & \multicolumn{1}{c|}{1.41}          & 12               & \multicolumn{1}{c|}{1.19}          & 11               & \multicolumn{1}{c|}{8.33}                  & 11                \\ \hline
Memory bug                & \multicolumn{1}{c|}{8.00}          & 1                & \multicolumn{1}{c|}{3.19}          & 2                & \multicolumn{1}{c|}{3.00}          & 3                & \multicolumn{1}{c|}{2.00}                  & 1                \\ \hline
Performance bug           & \multicolumn{1}{c|}{1.50}          & 12               & \multicolumn{1}{c|}{2.27}          & 7                & \multicolumn{1}{c|}{-}             & -                & \multicolumn{1}{c|}{9.50}                  & 12                \\ \hline
Processor bug             & \multicolumn{1}{c|}{2.60}          & 8                & \multicolumn{1}{c|}{2.67}          & 6                & \multicolumn{1}{c|}{3.15}          & 2                & \multicolumn{1}{c|}{5.33}                  & 2                \\ \hline
Test bug                  & \multicolumn{1}{c|}{3.16}          & 3                & \multicolumn{1}{c|}{1.78}          & 10               & \multicolumn{1}{c|}{2.04}          & 8                & \multicolumn{1}{c|}{7.00}                  & 9=                \\ \hline
Version compatibility bug & \multicolumn{1}{c|}{2.00}          & 11               & \multicolumn{1}{c|}{2.25}          & 8                & \multicolumn{1}{c|}{3.17}          & 1                & \multicolumn{1}{c|}{6.67}                  & 6=                \\ \hline
Visualization bug         & \multicolumn{1}{c|}{2.78}          & 6                & \multicolumn{1}{c|}{1.67}          & 11               & \multicolumn{1}{c|}{2.36}          & 4                & \multicolumn{1}{c|}{7.00}                  & 9=                \\ \hline
\end{tabular}
}
\label{table:NOFM}
\end{table*}

\textbf{(3) Entropy.} The average entropy and its ranking of each bug type in the three DLFs are shown in Table \ref{table:Entropy}. For MXNet, processor bugs and memory bugs are ranked in the top and bottom respectively; for PyTorch, deployment bugs and documentation bugs are ranked in the top and bottom respectively; and for TensorFlow, data bugs and documentation bugs are ranked in the top and bottom respectively. Taking the three DLFs as a whole, deployment bugs, data bugs, and processor bugs are the top 3 bug types that have the most entropy during bug fixing; build bugs, documentation bugs, and visualization bugs have the least entropy during bug fixing.

\begin{table*}[h]
\centering
\caption{Average Entropy and Its Ranking of Each Bug Type in DLFs (RQ2).} %
\scalebox{0.85}{
\begin{tabular}{|c|cc|cc|cc|cc|}
\hline
\multicolumn{1}{|l|}{}    & \multicolumn{2}{c|}{\textbf{MXNet}}                      & \multicolumn{2}{c|}{\textbf{PyTorch}}                    & \multicolumn{2}{c|}{\textbf{TensorFlow}}                 & \multicolumn{2}{c|}{\textbf{Total}}                           \\ \hline
\textbf{Bug type}         & \multicolumn{1}{c|}{\textbf{Entropy}} & \textbf{Ranking} & \multicolumn{1}{c|}{\textbf{Entropy}} & \textbf{Ranking} & \multicolumn{1}{c|}{\textbf{Entropy}} & \textbf{Ranking} & \multicolumn{1}{c|}{\textbf{Mean ranking}} & \textbf{InteRanking} \\ \hline
Algorithm design bug      & \multicolumn{1}{c|}{0.65}             & 3                & \multicolumn{1}{c|}{0.40}             & 4                & \multicolumn{1}{c|}{0.61}             & 4                & \multicolumn{1}{c|}{3.67}                  & 4                \\ \hline
Build bug                 & \multicolumn{1}{c|}{0.37}             & 11               & \multicolumn{1}{c|}{0.18}             & 11               & \multicolumn{1}{c|}{0.24}             & 10               & \multicolumn{1}{c|}{10.67}                 & 11=               \\ \hline
Code bug                  & \multicolumn{1}{c|}{0.57}             & 5                & \multicolumn{1}{c|}{0.37}             & 6                & \multicolumn{1}{c|}{0.36}             & 8                & \multicolumn{1}{c|}{6.33}                  & 6                \\ \hline
Data bug                  & \multicolumn{1}{c|}{0.62}             & 4                & \multicolumn{1}{c|}{0.51}             & 2                & \multicolumn{1}{c|}{0.67}             & 1                & \multicolumn{1}{c|}{2.33}                  & 2                \\ \hline
Deployment bug            & \multicolumn{1}{c|}{0.66}             & 2                & \multicolumn{1}{c|}{0.56}             & 1                & \multicolumn{1}{c|}{0.62}             & 3                & \multicolumn{1}{c|}{2.00}                  & 1                \\ \hline
Documentation bug         & \multicolumn{1}{c|}{0.42}             & 9                & \multicolumn{1}{c|}{0.15}             & 12               & \multicolumn{1}{c|}{0.06}             & 11               & \multicolumn{1}{c|}{10.67}                 & 11=               \\ \hline
Memory bug                & \multicolumn{1}{c|}{0.25}             & 12               & \multicolumn{1}{c|}{0.33}             & 8                & \multicolumn{1}{c|}{0.64}             & 2                & \multicolumn{1}{c|}{7.33}                  & 7                \\ \hline
Performance bug           & \multicolumn{1}{c|}{0.44}             & 8                & \multicolumn{1}{c|}{0.34}             & 7                & \multicolumn{1}{c|}{-}                & -                & \multicolumn{1}{c|}{7.50}                  & 8                \\ \hline
Processor bug             & \multicolumn{1}{c|}{0.80}             & 1                & \multicolumn{1}{c|}{0.46}             & 3                & \multicolumn{1}{c|}{0.54}             & 6                & \multicolumn{1}{c|}{3.33}                  & 3                \\ \hline
Test bug                  & \multicolumn{1}{c|}{0.46}             & 7                & \multicolumn{1}{c|}{0.25}             & 9                & \multicolumn{1}{c|}{0.39}             & 7                & \multicolumn{1}{c|}{7.67}                  & 9                \\ \hline
Version compatibility bug & \multicolumn{1}{c|}{0.46}             & 6                & \multicolumn{1}{c|}{0.37}             & 5                & \multicolumn{1}{c|}{0.58}             & 5                & \multicolumn{1}{c|}{5.33}                  & 5                \\ \hline
Visualization bug         & \multicolumn{1}{c|}{0.39}             & 10               & \multicolumn{1}{c|}{0.25}             & 10               & \multicolumn{1}{c|}{0.32}             & 9                & \multicolumn{1}{c|}{9.67}                  & 10               \\ \hline
\end{tabular}
}
\label{table:Entropy}
\end{table*}

\subsubsection{Complexity of Communication during Bug Fixing}
We present the results regarding the impact of different types of bugs on the complexity of communication in terms of NODP and NOC of bug-fixing pull requests in the three DLFs.

\textbf{(1) NODP.} The average NODP and its ranking of each bug type in the three DLFs are shown in Table \ref{table:NOUP}. For MXNet, documentation bugs and performance bugs are ranked in the top and bottom respectively; for PyTorch, memory bugs and documentation bugs are ranked in the top and bottom respectively; and for TensorFlow, memory bugs and visualization bugs are ranked in the top and bottom respectively. 
Taking the three DLFs as a whole, memory bugs, deployment bugs, and processor bugs are the top 3 bug types that have the most NODP during bug fixing; performance bugs, test bugs, and code bugs need the least NODP during bug fixing.
%Table \ref{table:NOUP} shows the relationship between bug types and NOUP. For MXNet, the five categories with greater impact are Documentation bug, Memory bug, Deployment bug, Visualization bug, and Processor bug; For PyTorch, Memory bug, Deployment bug, Algorithm design bug, Data bug, and Processor bug have a great impact; For TensorFlow, they are Memory bug, Processor bug, Version compatibility bug, Algorithm design bug, and Build bug.

\begin{table*}[h]
\centering
\caption{Average NODP and Its Ranking of Each Bug Type in DLFs  (RQ2).} %
\scalebox{0.85}{
\begin{tabular}{|c|cc|cc|cc|cc|}
\hline
\multicolumn{1}{|l|}{}    & \multicolumn{2}{c|}{\textbf{MXNet}}                   & \multicolumn{2}{c|}{\textbf{PyTorch}}                 & \multicolumn{2}{c|}{\textbf{TensorFlow}}              & \multicolumn{2}{c|}{\textbf{Total}}                           \\ \hline
\textbf{Bug type}         & \multicolumn{1}{c|}{\textbf{NODP}} & \textbf{Ranking} & \multicolumn{1}{c|}{\textbf{NODP}} & \textbf{Ranking} & \multicolumn{1}{c|}{\textbf{NODP}} & \textbf{Ranking} & \multicolumn{1}{c|}{\textbf{Mean ranking}} & \textbf{InteRanking} \\ \hline
Algorithm design bug      & \multicolumn{1}{c|}{4.05}          & 7                & \multicolumn{1}{c|}{5.53}          & 3                & \multicolumn{1}{c|}{6.18}          & 4                & \multicolumn{1}{c|}{4.67}                  & 4                \\ \hline
Build bug                 & \multicolumn{1}{c|}{4.06}          & 6                & \multicolumn{1}{c|}{5.13}          & 7                & \multicolumn{1}{c|}{6.13}          & 5                & \multicolumn{1}{c|}{6.00}                  & 5                \\ \hline
Code bug                  & \multicolumn{1}{c|}{3.93}          & 9                & \multicolumn{1}{c|}{4.96}          & 10               & \multicolumn{1}{c|}{5.91}          & 8                & \multicolumn{1}{c|}{9.00}                  & 10               \\ \hline
Data bug                  & \multicolumn{1}{c|}{3.89}          & 10               & \multicolumn{1}{c|}{5.42}          & 4                & \multicolumn{1}{c|}{6.06}          & 7                & \multicolumn{1}{c|}{7.00}                  & 7                \\ \hline
Deployment bug            & \multicolumn{1}{c|}{4.57}          & 3                & \multicolumn{1}{c|}{5.58}          & 2                & \multicolumn{1}{c|}{6.12}          & 6                & \multicolumn{1}{c|}{3.67}                  & 2                \\ \hline
Documentation bug         & \multicolumn{1}{c|}{4.80}          & 1                & \multicolumn{1}{c|}{4.62}          & 12               & \multicolumn{1}{c|}{5.83}          & 9                & \multicolumn{1}{c|}{7.33}                  & 9                \\ \hline
Memory bug                & \multicolumn{1}{c|}{4.75}          & 2                & \multicolumn{1}{c|}{5.85}          & 1                & \multicolumn{1}{c|}{8.17}          & 1                & \multicolumn{1}{c|}{1.33}                  & 1                \\ \hline
Performance bug           & \multicolumn{1}{c|}{3.00}          & 12               & \multicolumn{1}{c|}{4.82}          & 11               & \multicolumn{1}{c|}{-}             & -                & \multicolumn{1}{c|}{11.50}                 & 12               \\ \hline
Processor bug             & \multicolumn{1}{c|}{4.10}          & 5                & \multicolumn{1}{c|}{5.36}          & 5                & \multicolumn{1}{c|}{6.62}          & 2                & \multicolumn{1}{c|}{4.00}                  & 3                \\ \hline
Test bug                  & \multicolumn{1}{c|}{3.79}          & 11               & \multicolumn{1}{c|}{5.01}          & 8                & \multicolumn{1}{c|}{5.65}          & 10               & \multicolumn{1}{c|}{9.67}                  & 11               \\ \hline
Version compatibility bug & \multicolumn{1}{c|}{4.00}          & 8                & \multicolumn{1}{c|}{4.96}          & 9                & \multicolumn{1}{c|}{6.33}          & 3                & \multicolumn{1}{c|}{6.67}                  & 6                \\ \hline
Visualization bug         & \multicolumn{1}{c|}{4.11}          & 4                & \multicolumn{1}{c|}{5.17}          & 6                & \multicolumn{1}{c|}{5.00}          & 11               & \multicolumn{1}{c|}{7.00}                  & 7                \\ \hline

\end{tabular}
}
\label{table:NOUP}
\end{table*}

\begin{table*}[h]
\centering
\caption{Average NOC and Its Ranking of Each Bug Type in DLFs  (RQ2).} %
\scalebox{0.85}{
\begin{tabular}{|c|cc|cc|cc|cc|}
\hline
\multicolumn{1}{|l|}{}    & \multicolumn{2}{c|}{\textbf{MXNet}}                  & \multicolumn{2}{c|}{\textbf{PyTorch}}                & \multicolumn{2}{c|}{\textbf{TensorFlow}}             & \multicolumn{2}{c|}{\textbf{Total}}                           \\ \hline
\textbf{Bug type}         & \multicolumn{1}{c|}{\textbf{NOC}} & \textbf{Ranking} & \multicolumn{1}{c|}{\textbf{NOC}} & \textbf{Ranking} & \multicolumn{1}{c|}{\textbf{NOC}} & \textbf{Ranking} & \multicolumn{1}{c|}{\textbf{Mean ranking}} & \textbf{InteRanking} \\ \hline
Algorithm design bug      & \multicolumn{1}{c|}{4.00}         & 8                & \multicolumn{1}{c|}{6.26}         & 4                & \multicolumn{1}{c|}{4.09}         & 3                & \multicolumn{1}{c|}{5.00}                  & 3                \\ \hline
Build bug                 & \multicolumn{1}{c|}{4.03}         & 7                & \multicolumn{1}{c|}{5.20}         & 7                & \multicolumn{1}{c|}{3.55}         & 4                & \multicolumn{1}{c|}{6.00}                  & 5                \\ \hline
Code bug                  & \multicolumn{1}{c|}{3.87}         & 9                & \multicolumn{1}{c|}{5.15}         & 8                & \multicolumn{1}{c|}{3.22}         & 6                & \multicolumn{1}{c|}{7.67}                  & 9=              \\ \hline
Data bug                  & \multicolumn{1}{c|}{3.51}         & 10               & \multicolumn{1}{c|}{6.37}         & 3                & \multicolumn{1}{c|}{2.90}         & 8                & \multicolumn{1}{c|}{7.00}                  & 6=               \\ \hline
Deployment bug            & \multicolumn{1}{c|}{4.57}         & 6                & \multicolumn{1}{c|}{6.13}         & 5                & \multicolumn{1}{c|}{3.24}         & 5                & \multicolumn{1}{c|}{5.33}                  & 4                \\ \hline
Documentation bug         & \multicolumn{1}{c|}{4.60}         & 5                & \multicolumn{1}{c|}{3.97}         & 12               & \multicolumn{1}{c|}{2.23}         & 10               & \multicolumn{1}{c|}{9.00}                  & 11               \\ \hline
Memory bug                & \multicolumn{1}{c|}{7.75}         & 1                & \multicolumn{1}{c|}{10.74}        & 1                & \multicolumn{1}{c|}{8.67}         & 1                & \multicolumn{1}{c|}{1.00}                  & 1                \\ \hline
Performance bug           & \multicolumn{1}{c|}{2.00}         & 12               & \multicolumn{1}{c|}{6.86}         & 2                & \multicolumn{1}{c|}{-}            & -                & \multicolumn{1}{c|}{7.00}                  & 6=               \\ \hline
Processor bug             & \multicolumn{1}{c|}{6.30}         & 3                & \multicolumn{1}{c|}{5.56}         & 6                & \multicolumn{1}{c|}{4.77}         & 2                & \multicolumn{1}{c|}{3.67}                  & 2                \\ \hline
Test bug                  & \multicolumn{1}{c|}{5.32}         & 4                & \multicolumn{1}{c|}{4.99}         & 9                & \multicolumn{1}{c|}{2.30}         & 9                & \multicolumn{1}{c|}{7.33}                  & 8                \\ \hline
Version compatibility bug & \multicolumn{1}{c|}{2.70}         & 11               & \multicolumn{1}{c|}{4.50}         & 11               & \multicolumn{1}{c|}{3.17}         & 7                & \multicolumn{1}{c|}{9.67}                  & 12               \\ \hline
Visualization bug         & \multicolumn{1}{c|}{7.22}         & 2                & \multicolumn{1}{c|}{4.83}         & 10               & \multicolumn{1}{c|}{2.18}         & 11               & \multicolumn{1}{c|}{7.67}                  & 9=              \\ \hline

\end{tabular}
}
\label{table:NOC}
\end{table*}

%\textbf{In terms of the relationship between bug types and NOUP, Memory bug and Processor bug have a strong impact on the three frameworks; Algorithm design bug and Deployment bug have strong influence on two of the three frameworks.}

%Table \ref{table:NOC} shows the relationship between bug types and NOC. For MXNet, the five categories with greater impact are Memory bug, Visualization bug, Processor bug, Test bug, and Documentation bug; For PyTorch, Memory bug, Performance bug, Data bug, Algorithm design bug, and Deployment bug have a great impact; For TensorFlow, they are Memory bug, Processor bug, Algorithm design bug, Build bug, and Code bug.

%\textbf{In terms of the relationship between bug types and NOC, Memory bug has a strong impact on the three frameworks; Algorithm design bug and Processor bug have strong influence on two of the three frameworks.}

\textbf{(2) NOC.} The average NOC and its ranking of each bug type in the three DLFs are shown in Table \ref{table:NOUP}. For MXNet, memory bugs and performance bugs are ranked in the top and bottom respectively; for PyTorch, memory bugs and documentation bugs are ranked in the top and bottom respectively; and for TensorFlow, memory bugs and version compatibility  bugs are ranked in the top and bottom respectively. 
Taking the three DLFs as a whole, memory, processor, and algorithm design bugs are the top 3 bug types that have the most NODP during bug fixing; version compatibility, documentation, code, and visualization bugs have the least NOC during bug fixing.

\textbf{Answer to RQ2:} 
%For all the bug types, \textit{deployment bugs} are ranked on the top in terms of open time; memory bugs are ranked on the top with respect to the code change complexity of bug fixes in terms of LOCM and NOFM, and deployment bugs are ranked on the top with respect to the code change complexity of bug fixes in terms of entropy; 
\textit{Deployment bugs} negatively impact the development of DLFs the most in terms of open time; \textit{deployment bugs} negatively impact code change complexity the most in terms of entropy; \textit{memory bugs} negatively impact code change complexity the most in terms of LOCM and NOFM; and \textit{memory bugs} negatively impact communication complexity the most in terms of NODP and NOC.

%\begin{tcolorbox}[colback=white!5]
%  \textbf{Answer to RQ2:} \red{We found that various types of bugs will show different results in the metrics, but there will also be similarities. To sum up, we found that Algorithm design bug, Deployment bug, Memory bug and Processor bug show higher fix complexity in the results. Relatively speaking, Build bug, Documentation bug, Test bug and Visualization bug will present relatively low fix complexity.}
%\end{tcolorbox}

\subsection{Proportions and Distribution of MPL Bug Fixes (RQ3)}
\textbf{Proportions of MPL bugs.} 
As shown in TABLE \ref{table:distributionMPL}, among the 189 bugs analyzed in MXNet, 54 are MPL bugs (which bug-fixing pull requests involve multiple PLs), accounting for 28.6\%. In PyTorch, 291 out of the analyzed 926 bugs are MPL bugs, accounting for 31.4\%. In TensorFlow, 61 out of the 382 analyzed bugs are MPL bugs, accounting for 16.0\%.

\textbf{Distribution of MPL bugs over bug types.} 
As shown in TABLE \ref{table:distributionMPL}, in MXNet, algorithm design bugs are the top bug type that has the largest proportions of MPL bugs, while deployment bugs, performance bugs, and visualization bugs are the three bug types that have no MPL bugs; in PyTorch, data bugs are the top bug type that has the largest proportions of MPL bugs, while visualization bugs are the only bug type that has no MPL bugs; In TensorFlow, processor bugs are the top bug type that has the largest proportions of MPL bugs, while version compatibility bugs are the only bug type that have no MPL bugs. Performance bugs in TensorFlow are not discussed here because there are no performance bugs.

\begin{table*}[h]
\centering
\caption{Type Distribution of Bugs with MPL and SPL Fixes (RQ3).} %
\scalebox{0.85}{
\begin{tabular}{|c|ccc|ccc|ccc|}
\hline
\multirow{2}{*}{\textbf{}} & \multicolumn{3}{c|}{\textbf{MXNet}}                                                         & \multicolumn{3}{c|}{\textbf{PyTorch}}                                                       & \multicolumn{3}{c|}{\textbf{TensorFlow}}                                                    \\ \cline{2-10} 
                           & \multicolumn{1}{c|}{\textbf{SPL}} & \multicolumn{1}{c|}{\textbf{MPL}} & \textbf{\%MPL} & \multicolumn{1}{c|}{\textbf{SPL}} & \multicolumn{1}{c|}{\textbf{MPL}} & \textbf{\%MPL} & \multicolumn{1}{c|}{\textbf{SPL}} & \multicolumn{1}{c|}{\textbf{MPL}} & \textbf{\%MPL} \\ \hline
Algorithm design bug       & \multicolumn{1}{c|}{10}           & \multicolumn{1}{c|}{11}           & 52.4             & \multicolumn{1}{c|}{68}           & \multicolumn{1}{c|}{51}           & 42.9             & \multicolumn{1}{c|}{30}           & \multicolumn{1}{c|}{4}            & 11.8             \\ \hline
Build bug                  & \multicolumn{1}{c|}{29}           & \multicolumn{1}{c|}{3}            & 9.4              & \multicolumn{1}{c|}{106}          & \multicolumn{1}{c|}{6}            & 5.4              & \multicolumn{1}{c|}{31}           & \multicolumn{1}{c|}{0}            & 0.0              \\ \hline
Code bug                   & \multicolumn{1}{c|}{12}           & \multicolumn{1}{c|}{3}            & 20.0             & \multicolumn{1}{c|}{67}           & \multicolumn{1}{c|}{14}           & 17.3             & \multicolumn{1}{c|}{37}           & \multicolumn{1}{c|}{8}            & 17.8             \\ \hline
Data bug                   & \multicolumn{1}{c|}{23}           & \multicolumn{1}{c|}{22}           & 48.9             & \multicolumn{1}{c|}{72}           & \multicolumn{1}{c|}{83}           & 53.6             & \multicolumn{1}{c|}{62}           & \multicolumn{1}{c|}{34}           & 35.4             \\ \hline
Deployment bug             & \multicolumn{1}{c|}{7}            & \multicolumn{1}{c|}{0}            & 0.0              & \multicolumn{1}{c|}{63}           & \multicolumn{1}{c|}{43}           & 40.6             & \multicolumn{1}{c|}{23}           & \multicolumn{1}{c|}{2}            & 8.0              \\ \hline
Documentation bug          & \multicolumn{1}{c|}{12}           & \multicolumn{1}{c|}{3}            & 20.0             & \multicolumn{1}{c|}{27}           & \multicolumn{1}{c|}{2}            & 6.9              & \multicolumn{1}{c|}{78}           & \multicolumn{1}{c|}{2}            & 2.5              \\ \hline
Memory bug                 & \multicolumn{1}{c|}{2}            & \multicolumn{1}{c|}{2}            & 50.0             & \multicolumn{1}{c|}{18}           & \multicolumn{1}{c|}{9}            & 33.3             & \multicolumn{1}{c|}{5}            & \multicolumn{1}{c|}{1}            & 16.7             \\ \hline
Performance bug            & \multicolumn{1}{c|}{2}            & \multicolumn{1}{c|}{0}            & 0.0              & \multicolumn{1}{c|}{17}           & \multicolumn{1}{c|}{5}            & 22.7             & \multicolumn{1}{c|}{-}            & \multicolumn{1}{c|}{-}            & -                   \\ \hline
Processor bug              & \multicolumn{1}{c|}{5}            & \multicolumn{1}{c|}{5}            & 50.0             & \multicolumn{1}{c|}{73}           & \multicolumn{1}{c|}{59}           & 44.7             & \multicolumn{1}{c|}{6}            & \multicolumn{1}{c|}{7}            & 53.9             \\ \hline
Test bug                   & \multicolumn{1}{c|}{17}           & \multicolumn{1}{c|}{2}            & 10.5             & \multicolumn{1}{c|}{75}           & \multicolumn{1}{c|}{8}            & 9.6              & \multicolumn{1}{c|}{21}           & \multicolumn{1}{c|}{2}            & 8.7              \\ \hline
Version compatibility bug  & \multicolumn{1}{c|}{7}            & \multicolumn{1}{c|}{3}            & 30.0             & \multicolumn{1}{c|}{37}           & \multicolumn{1}{c|}{11}           & 22.9             & \multicolumn{1}{c|}{18}           & \multicolumn{1}{c|}{0}            & 0.0              \\ \hline
Visualization bug          & \multicolumn{1}{c|}{9}            & \multicolumn{1}{c|}{0}            & 0.0              & \multicolumn{1}{c|}{12}           & \multicolumn{1}{c|}{0}            & 0.0              & \multicolumn{1}{c|}{10}           & \multicolumn{1}{c|}{1}            & 9.1              \\ \hline
\textbf{All bug types}              & \multicolumn{1}{c|}{\textbf{135}}          & \multicolumn{1}{c|}{\textbf{54}}           & \textbf{28.6}             & \multicolumn{1}{c|}{\textbf{635}}          & \multicolumn{1}{c|}{\textbf{291}}          & \textbf{31.4}             & \multicolumn{1}{c|}{\textbf{321}}          & \multicolumn{1}{c|}{\textbf{61}}           & \textbf{16.0}             \\ \hline

\end{tabular}
}
\label{table:distributionMPL}
\end{table*}

\textbf{Use of PL combinations in MPL bug fixes.} The results of MPL bugs over different PL combinations are shown in TABLE \ref{table:PLCombination}. In MXNet, only the combination of Python and C/C++ is used in all the MPL bug fixes. In PyTorch, the combination of Python and C/C++ is used in 268 (92.1\%) MPL bug fixes, and the combination of Python and Objective-C is used in 17 (5.8\%) MPL bug fixes, the combination of C/C++ and Objective-C is used in 3 MPL bug fixes, the combination of Python, C/C++, and Objective-C is used in 2 MPL bug fixes, and the combination of C/C++, Ruby, and Objective-C is used in 1 MPL bug fix. In TensorFlow, the combination of Python and C/C++ is used in 60 (98.4\%) MPL bug fixes, and only one MPL bug fix uses the combination of Python and Go.

\textbf{Answer to RQ3:} 28.6\%, 31.4\%, and 16.0\% bug fixes are MPL fixes in MXNet, PyTorch, and TensorFlow, respectively. \textit{Algorithm design bugs}, \textit{data bugs}, and \textit{processor bugs} are the bug types that hold the largest proportions in MXNet, PyTorch, and TensorFlow, respectively. The PL combination of Python and C/C++ is most popular in MPL bug fixes of the DLFs.

%\begin{tcolorbox}[colback=white!5]
%  \textbf{Answer to RQ3:} In MXNet, 28.57\%, 31.42\%, and 15.97\% bug fixes are MPL fixes in MXNet, PyTorch, and TensorFlow, respectively. \red{XXX.} The PL combination of Python and C/C++ is most popular in MPL bug fixes of the three DLFs.
%\end{tcolorbox}

\begin{table}[h]
\centering
\caption{PL Combinations of MPL Bug Fixes (RQ3).}
\scalebox{0.85}{
\begin{tabular}{|c|cc|cc|cc|}
\hline
\multicolumn{1}{|l|}{\textbf{}} & \multicolumn{2}{c|}{\textbf{MXNet}}            & \multicolumn{2}{c|}{\textbf{PyTorch}}          & \multicolumn{2}{c|}{\textbf{TensorFlow}}       \\ \hline
\textbf{PL combination}         & \multicolumn{1}{c|}{\textbf{\#}} & \textbf{\%} & \multicolumn{1}{c|}{\textbf{\#}} & \textbf{\%} & \multicolumn{1}{c|}{\textbf{\#}} & \textbf{\%} \\ \hline
C/C++, Python                   & \multicolumn{1}{c|}{54}          & 100.0      & \multicolumn{1}{c|}{268}         & 92.1       & \multicolumn{1}{c|}{60}          & 98.4       \\ \hline
Objective-C, Python             & \multicolumn{1}{c|}{0}           & 0.0        & \multicolumn{1}{c|}{17}          & 5.8        & \multicolumn{1}{c|}{0}           & 0.0        \\ \hline
C/C++, Objective-C              & \multicolumn{1}{c|}{0}           & 0.0        & \multicolumn{1}{c|}{3}           & 1.0        & \multicolumn{1}{c|}{0}           & 0.0        \\ \hline
Go, Python                      & \multicolumn{1}{c|}{0}           & 0.0        & \multicolumn{1}{c|}{0}           & 0.0        & \multicolumn{1}{c|}{1}           & 1.6        \\ \hline
C/C++, Objective-C, Python      & \multicolumn{1}{c|}{0}           & 0.0        & \multicolumn{1}{c|}{2}           & 0.7        & \multicolumn{1}{c|}{0}           & 0.0        \\ \hline
C/C++, Objective-C, Ruby        & \multicolumn{1}{c|}{0}           & 0.0        & \multicolumn{1}{c|}{1}           & 0.3        & \multicolumn{1}{c|}{0}           & 0.0        \\ \hline
\end{tabular}
}
\label{table:PLCombination}
\end{table}

\subsection{Impact of the Use of Multiple PLs on Bug Fixing (RQ4)}
We further compared the six impact indicators of MPL bug fixes with those of SPL bug fixes in the DLFs, in order to understand whether there are significant differences on the characteristics between MPL and SPL bug fixes. The results are shown in TABLE \ref{table:impactMPL}. (1) In MXNet, the LOCM and NOFM of MPL bug fixes are significantly larger than the LOCM and NOFM of SPL bug fixes, respectively, while there are no significant differences between MPL bug fixes and SPL bug fixes on OT, entropy, NODP, and NOC. (2) In PyTorch, every indicator of MPL bug fixes is significantly greater than that of SPL bug fixes. (3) In TensorFlow, the LOCM, NOFM, and Entropy of MPL bug fixes are significantly larger than those of SPL bug fixes, respectively, while there are no significant differences between MPL bug fixes and SPL bug fixes on OT, NODP, and NOC. 

\textbf{Answer to RQ4:} No impact indicators of MPL bug fixes are significantly smaller than those of SPL bug fixes. Code change complexity of MPL bug fixes is significantly greater than that of SPL bug fixes in terms of LOCM and NOFM.

\begin{table*}[h]
\centering
\caption{Mann-Whitney U Test results regarding impacts of the use of multiple PLs on the fixes of bugs (RQ4).}
\scalebox{0.85}{
\begin{tabular}{|c|ccc|ccc|ccc|}
\hline
\multicolumn{1}{|l|}{\textbf{}} & \multicolumn{3}{c|}{\textbf{MXNet}}                                                                   & \multicolumn{3}{c|}{\textbf{PyTorch}}                                                    & \multicolumn{3}{c|}{\textbf{TensorFlow}}                                                              \\ \hline
\multicolumn{1}{|l|}{\textbf{}} & \multicolumn{1}{c|}{\textbf{SPL bugs}} & \multicolumn{1}{c|}{\textbf{MPL bugs}} & \textbf{\textit{p-value}}              & \multicolumn{1}{c|}{\textbf{SPL bugs}} & \multicolumn{1}{c|}{\textbf{MPL bugs}} & \textbf{\textit{p-value}} & \multicolumn{1}{c|}{\textbf{SPL bugs}} & \multicolumn{1}{c|}{\textbf{MPL bugs}} & \textbf{\textit{p-value}}              \\ \hline
OT                              & \multicolumn{1}{c|}{51.51}        & \multicolumn{1}{c|}{47.01}        & \cellcolor[HTML]{D3D1D1}0.643 & \multicolumn{1}{c|}{48.21}        & \multicolumn{1}{c|}{59.24}        & \textless{}0.001 & \multicolumn{1}{c|}{73.04}        & \multicolumn{1}{c|}{87.00}        & \cellcolor[HTML]{D3D1D1}0.227 \\ \hline
LOCM                            & \multicolumn{1}{c|}{35.01}        & \multicolumn{1}{c|}{57.63}        & \textless{}0.001              & \multicolumn{1}{c|}{34.37}        & \multicolumn{1}{c|}{76.36}        & \textless{}0.001 & \multicolumn{1}{c|}{27.69}        & \multicolumn{1}{c|}{36.80}        & \textless{}0.001              \\ \hline
NOFM                            & \multicolumn{1}{c|}{2.25}         & \multicolumn{1}{c|}{4.41}         & \textless{}0.001              & \multicolumn{1}{c|}{2.07}         & \multicolumn{1}{c|}{3.95}         & \textless{}0.001 & \multicolumn{1}{c|}{1.89}         & \multicolumn{1}{c|}{2.67}         & \textless{}0.001              \\ \hline
Entropy                         & \multicolumn{1}{c|}{0.53}         & \multicolumn{1}{c|}{0.71}         & \cellcolor[HTML]{D3D1D1}0.059 & \multicolumn{1}{c|}{0.34}         & \multicolumn{1}{c|}{0.57}         & \textless{}0.001 & \multicolumn{1}{c|}{0.39}         & \multicolumn{1}{c|}{0.80}         & \textless{}0.001              \\ \hline
NODP                            & \multicolumn{1}{c|}{3.93}         & \multicolumn{1}{c|}{4.31}         & \cellcolor[HTML]{D3D1D1}0.479 & \multicolumn{1}{c|}{5.16}         & \multicolumn{1}{c|}{5.62}         & \textless{}0.001 & \multicolumn{1}{c|}{6.13}         & \multicolumn{1}{c|}{5.93}         & \cellcolor[HTML]{D3D1D1}0.267 \\ \hline
NOC                             & \multicolumn{1}{c|}{3.91}         & \multicolumn{1}{c|}{5.07}         & \cellcolor[HTML]{D3D1D1}0.057 & \multicolumn{1}{c|}{5.44}         & \multicolumn{1}{c|}{6.96}         & \textless{}0.001 & \multicolumn{1}{c|}{3.30}         & \multicolumn{1}{c|}{2.80}         & \cellcolor[HTML]{D3D1D1}0.837 \\ \hline

\end{tabular}
}
\label{table:impactMPL}
\end{table*}

%\begin{tcolorbox}[colback=white!5]
%  \textbf{Answer to RQ4:} No impact indicators of MPL bug fixes are significantly smaller than those of SPL bug fixes. Code change complexity of MPL bug fixes is significantly greater than that of SPL bug fixes in terms of LOCM and NOFM.
%\end{tcolorbox}

\section{Discussion}
\label{chap:discus}

%In this section, we interpret the results of the study according to the RQs and discuss the implications of the results for both practitioners and researchers. 

\subsection{Interpretation of Study Results}
\emph{\textbf{RQ1}}: 
%From the perspective of proportion, the types of bugs that account for a high proportion in MXNet are Data bug, Build bug, Algorithm design bug, and Test bug; In PyTorch, Data bug, Processor bug, Algorithm design bug, and Build bug account for a high proportion of bug types; In TensorFlow, Data bug, Document bug, Code bug, and Algorithm design bug account for a high proportion, as shown in Figure \ref{fig:BugTypes}.
(1) Data bugs and algorithm design bugs are very common in these algorithm intensive software systems. The most common causes are problems in the input and output links of operators or neural network layers, e.g., the lack of input type checks (\textit{TensorFlow \#13506}), the lack of necessary edge checks (\textit{PyTorch \#1653}), and the output of unexpected results (\textit{MXNet \#8303}). 
%\red{Therefore, when developers design models or custom operators, it is necessary to conduct a complete test, such as Zhang \cite{zhang2020detecting} proposed a static analysis method based on abstract interpretation to detect digital bugs in neural architecture.}
(2) Documentation bugs account for a relatively large proportion (20.94\%) in TensorFlow, which indicates that during the development of TensorFlow, bugs in updating documents is often a major difficulty for developers. This was confirmed by the results of our searches for the keyword "document" in Stack Overflow regarding the three DLFs: 13 hits returned for MXNet, 174 hits returned for PyTorch, 913 hits returned for TensorFlow. 
%Although TensorFlow was released earlier than the first two frameworks, such a big difference may also indicate that the documents of TensorFlow are less mature than MXNet and PyTorch.
(3) In PyTorch, processor bugs are also a major bug type, which resonates the great efforts of PyTorch developers spent in data synchronization and parallel and multiprocessing \cite{r17}. 
(4) The percentage of performance bugs in the three DLFs is very small. We checked the issues labeled with "performance" in the three DLFs and found that most of such issues are not considered as bugs by developers. 
%Whether this performance related problem should be considered as a bug or a problem worth discussing. It is not a bug in performance, but it does affect the quality attributes of the system.

\emph{\textbf{RQ2}}: 
%The development difficulty can be indicated by the open time of bugs, the code change complexity of bug fixes, and the communication complexity during bug fixing. 
(1) Deployment bugs are ranked on the top in the integrated ranking with respect to the OT and entropy of bugs, and memory bugs are ranked on the top in the integrated ranking with respect to the LOCM and NOFM of bugs. As evidenced in \cite{r29}, a larger code change complexity and longer OT often mean higher maintenance cost of a bug. In this sense, deployment bugs and memory bugs may incur much maintenance cost.   
(2) Memory bugs are ranked on the top in the integrated ranking with respect to the NODP and NOC during bug fixing, which means the communication of fixing memory bugs is most complex in the sense that most developers are involved and the bugs are most discussed. It also resonates with the finding that the code change complexity of memory bugs is the largest in terms of LOCM and NOFM.

\emph{\textbf{RQ3}}: 
(1) The proportion of MPL bugs in TensorFlow is obviously smaller than that in MXNet and PyTorch. One main potential reason is that the main PL in TensorFlow accounts for 63.1\% (shown in TABLE \ref{table:demographicInfo}), which is much larger than that in MXNet (48.6\%) and PyTorch (49.4\%). It implies that there is much less chances for the main PL to work with other PLs in TensorFlow.
(2) The combination of Python and C/C++ is the most used PL combination in fixing MPL bugs of the selected DLFs. This is also the PL choice of most popular DLFs. We speculate the reason is that Python is probably the most comfortable PL for many data scientists and machine learning experts, and it is easy to integrate C/C++ components and is compatible with a large number of computing libraries. They prefer to use C/C++ as the computational part to improve performance \cite{mu2019review}. We conducted a sampling analysis of MPL fixes to verify our conjecture that the vast majority of MPL bug fixes did occur in the back end. When a bug in the back end is resolved, developers need to test whether the issue is resolved in the front end. 

\emph{\textbf{RQ4}}: In MXNet, PyTorch, and TensorFlow, it is more difficult and costlier to fix MPL bugs than SPL bugs, in the sense that (1) MPL bug fixes have significantly larger code change complexity than SPL bug fixes in terms of LOCM and NOFM in the three DLFs, and (2) no impact indicators of MPL bug fixes are significantly smaller than those of SPL bug fixes in the three DLFs. This is consistent with the previous research results on MPL commits in Apache projects \cite{r29}, which suggests that the use of multiple PLs has a non-negligible impact on the development of DLFs.

%According to our research results, in MXNet, PyTorch and TensorFlow, compared with SPL fix, MPL fix will show a significant growth trend in OT, LOCM, NOFM, NOUP, NOC and Entropy, which indicates that MPL fix will have higher modification complexity. This is consistent with the previous research in the Apache project [20]. This proves that in the DLF, the selection of MPL will have a considerable impact on the software system.

\subsection{Implications for Researchers}
Firstly, we classified the bugs in the selected DLFs based on bug labels, which is greatly helpful for researchers to have a general understanding on the distribution of bugs in DLFs, thereby facilitating further investigation of these bugs. 
%Through our research results, researchers will know which bugs will cause relatively higher repair complexity to the system, so that we can conduct more in-depth research on them. 
%\blue{Secondly, considering a DLFs as a MPL software system, we discuss the universality of using multiple PLs to repair bugs in DLFs. This can help researchers understand the MPL bugs in the DLFs and the impact of this technology selection of MPL.}
%Secondly, it is worth further exploring how different combinations of PLs influence the development of DLFs, since the results of this exploration can guide the development to a more efficient way.
Secondly, it is worth further exploring how different combinations of PLs influence the development of DLFs, since the results of this exploration can guide the development to a more efficient way. 
Thirdly, researchers may put their efforts to study how the combination of C/C++ and Python may impact MPL bugs, because more than 92\% of MPL bugs involve this PL combination.
%\red{Finally, it is worth further exploring the causes of the differences on the open time and communication complexity of MPL bug fixes and SPL bug fixes between MXNet, TensorFlow and PyTorch.}
Finally, MPL bugs should receive more attention from researchers in the software engineering community but not limited to the DLF domain. Currently, MPL bugs are seldom explored. Due to the significant impact of MPL bugs, they need to be studied in more depth.

\subsection{Implications for Practitioners}
Firstly, our research results will enable developers to understand the distribution of bugs in the DLFs and which types of bugs are more efficient to solve first. We recommend developers to employ our bug classification of DLFs in practice thereby facilitating bug management by taking into account the impacts of different bug types. Besides, developers could choose more experienced developers to fix more complex bug types (e.g., \textit{memory bugs} and \textit{deployment bugs}) that can be identified through bug descriptions or related bug labels. For bugs with relatively low fix complexity, new developers can be assigned to fix them, thereby reducing the bug fixing cost. 
%Developers should fix the bug as soon as possible to reduce the extra interest generated by the bug. 
Meanwhile, the developers of DLFs usually leverage the advantages of different PLs to implement DLFs, however, the developers should be aware of that the use of multiple PLs for implementing DLFs is likely to increase the complexity of code changes, resulting in higher maintenance cost.

%\red{Therefore, when developers design models or custom operators, it is necessary to conduct a complete test, such as Zhang [36] proposed a static analysis method based on abstract interpretation to detect digital bugs in neural architecture.}

\section{Threats to Validity}
\label{chap:threats}

There are several threats to the validity of the study results. We discuss these threats according to the guidelines in \cite{RuHo2009}. 

\subsection{Construct Validity}
Construct validity is concerned with whether the values of the variables (listed in TABLE \ref{table:dataitem}) we obtained are consistent with the real values that we expected. A potential threat is that not all bugs resolved are explicitly linked to corresponding pull requests, which may negatively affect the representativeness of the collected bugs. Through our analysis, we confirmed that the bugs with explicit links to corresponding pull requests were not reported in a short time span and were not fixed by a small group of developers. Therefore, this threat is to some extent mitigated.
Another possible threat is that different developers may have biases during manual analysis, which may lead to inconsistent classification and incorrect bug tagging results. To reduce this threat, we adopted a four-step bug classification process, in which a pilot bug tagging process was adopted to identify and resolve possible disagreements with bug types (as described in Section \ref{DataAnalysisRQ1}).

%\subsection{Internal Validity}
%An internal threat to the effectiveness of our results may be the incorrect classification we made in the manual annotation process. In order to minimize the impact of this threat on our study results, we have three authors involved in the classification process. First of all, we fully understand the meaning of the labels developers have marked on bugs in advance, and then we get all the labels through the tool. We have classified these labels for the first round, but this round of classification is still unstable. In order to obtain more accurate classification results, we manually analyzed bugs through the problem itself with the help of grounded theory \cite{r4}. If a new classification or the original classification is inappropriate in the process of bug analysis, the first two authors will discuss to resolve the conflict and modify the original classification, and finally obtain a more stable classification result. Then we randomly selected 10\% of the bugs, and the second and third authors labeled and classified the bugs at the same time. We used Cohen's kappa \cite{r23} to measure the consistency between classifier. If the Kappa value is less than 90\%, three researchers will discuss to resolve the conflict, and then extract 10\% of the bug for analysis. We conducted two rounds of analysis until the kappa value exceeded 90\%.

\subsection{External Validity}
External validity is concerned with the generalizability of
the study results. A potential threat is whether the selected DLFs are representative enough. To select appropriate DLFs, we set a number of criteria as described in Section \ref{CaseSelection}, and the finally selected DLFs are MXNet, PyTorch, and TensorFlow. These three DLFs are very popular and support a wide range of DL applications. Thus, this threat is mitigated to some extent.
Another possible threat is that more than 92\% MPL bug fixes use the PL combination of Python and C/C++, which means significant imbalance regarding the use of multiple PLs. However, the PL combination of Python and C/C++ is natural choice in the development of DLFs by trading off the usability and performance of the two PLs.

%Another external threat to the effectiveness of our results is the threat from data. The first is the inaccuracy of the data. When we deeply studied the bug description and its pull request, we found that not all issues and pull requests in GitHub are strictly corresponding, and there are still the following situations: 1. Pull request corresponds to multiple issues; 2. Issue corresponds to multiple pull requests; 3. Pull request only solves partial issue; 4. Pull request is a fixed stack, which means that bug fix is only a part of the pull request (for example, PyTorch \# 83745). For case 1, if a pull request corresponds to multiple issues, it indicates that it has made greater efforts than a bug fix; For case 2, if one issue corresponds to multiple pull requests, we need to spend extra effort to review the code, but for the number of comments, reviews, and other indicators, simple addition is not a good choice; For cases three and four, the instability of the experiment will also be increased. We checked all pull requests and eliminated the bugs in the above four cases, which also reduced our dataset (from 1909 to 1497). In addition, the inaccuracy of the data statistical results is caused by the small amount of data. To reduce the impact of this error on the experimental results, we will discuss the three frameworks separately, and study the commonness of multiple groups of experimental results through multiple measurement indicators set, to obtain a relatively reliable result.

\subsection{Reliability}
Reliability is concerned with whether the same results can be obtained when the study is replicated by other researchers. A potential threat is related to the implementation of our software tool for data collection. The tool was implemented by the second author, and the code of the key functionalities had been reviewed by the first and third authors. In addition, sufficient tests were performed to ensure that the calculation of data items are correct. Thus, this threat had been alleviated. Another threat is related to the correctness of the Mann–Whitney U tests. Since only IBM SPSS was used to run the tests, this threat is minimized.

\section{Conclusions and Future Work}
\label{conclusions}
In this work, we conducted an empirical study on bugs in three mainstream DLFs (i.e., MXNet, PyTorch, and TensorFlow) and their impacts on the development of such DLFs. We collected 75114 issues and 92842 pull requests of the DLFs, and obtained 1497 bugs after a set of filterings. We manually analyzed the 1497 bugs, and got the following findings. 
(1) The bugs in DLFs can be classified into 12 types, e.g., \textit{algorithm design bugs}, \textit{data bugs}, and \textit{memory bugs}. Among the 12 bug types, \textit{data bugs} account for the largest proportion in all the three DLFs.
(2) \textit{Deployment bugs} and \textit{memory bugs} negatively impact the development of DLFs the most in various aspects.
(3) 28.6\%, 31.4\%, and 16.0\% of bugs in MXNet, PyTorch, and TensorFlow are MPL bugs respectively, and the PL combination of Python and C/C++ is most used in fixing more than 92\% MPL bugs in the three DLFs.
(4) The code change complexity of MPL bug fixes is significantly greater than that of SPL bug fixes in the three DLFs, while in PyTorch MPL bug fixes have longer open time and greater communication complexity.
%divided them into 12 categories according to their characteristics, and explored their symptoms and their impact on the project. At the same time, we explored the bug fixing in the DLF from the perspective of the MPL software system. We found that in the DLF, the selection of MPL will have a considerable global impact on the software system, and the bug fix complexity will be higher. With the continuous development of the DLF, more and more DLFs appear in people's vision. How to prevent, detect and solve bugs in the DLF is becoming increasingly important. 
In the future, we plan to expand the obtained dataset in this study, and construct bug prediction models for MPL bugs in DLFs. In addition, we are also interested in investigating MPL bugs in other application domains in depth.  
%In the future, we plan to expand the obtained dataset in this study, and automatically assign labels and priorities to issues through multi-classification algorithms, to better alert developers of DLFs which bugs should be more noteworthy.
\balance

\bibliographystyle{IEEEtran}

\bibliography{references}

%\printbibliography

\end{document}